\documentclass[twocolumn,a4paper,accepted=2022-11-24]{quantumarticle}
\pdfoutput=1

\usepackage{amsmath}
\usepackage{amssymb}
\usepackage{amsfonts}
\usepackage{mathrsfs}
\usepackage{bm, dsfont}
\usepackage{booktabs}
\usepackage{multirow}
\usepackage[usenames,dvipsnames]{color}
\usepackage{colortbl}
\usepackage[table]{xcolor}
\usepackage{enumitem}
\usepackage{dcolumn}
\usepackage{graphicx} 
\usepackage{epstopdf} 
\usepackage[numbers,sort&compress]{natbib}
\usepackage[colorlinks=true,linkcolor=blue,citecolor=blue,urlcolor=blue]{hyperref}
\usepackage{cleveref}
\usepackage{hypcap}
\usepackage{verbatim, float}
\usepackage{psfrag}
\usepackage[normalem]{ulem}
\usepackage{physics}		
\usepackage{siunitx}

\newcommand{\mean}[1]{\langle {#1} \rangle}
\newcommand{\prjct}[1]{\mathinner{|{#1}\rangle}\!\!\mathinner{\langle{#1}|}}

\newcommand{\id}{\mathds{1}}
\newcommand{\ii}{\mathrm{i}}

\renewcommand{\t}[1]{\mathrm{#1}}
\newcommand{\be}{\begin{equation}}
\newcommand{\ee}{\end{equation}}

\begin{document}

\title{Benchmarking single-photon sources \newline from an auto-correlation measurement}

\author{Pavel Sekatski}
\email{pavel.sekatski@unige.ch}
\affiliation{Department of Applied Physics, University of Geneva, Geneva, Switzerland}
\author{Enky Oudot}
\affiliation{ICFO - Institut de Ciencies Fotoniques, The Barcelona Institute
of Science and Technology, 08860 Castelldefels (Barcelona), Spain}
\author{Patrik Caspar}
\affiliation{Department of Applied Physics, University of Geneva, Geneva, Switzerland}
\author{Rob Thew}
\email{robert.thew@unige.ch}
\affiliation{Department of Applied Physics, University of Geneva, Geneva, Switzerland}
\author{Nicolas Sangouard}
\affiliation{Université Paris-Saclay, CEA, CNRS, Institut de physique théorique, 91191, Gif-sur-Yvette, France}

\begin{abstract}
Here we argue that the probability that a given source produces exactly a single photon is a natural quantity to benchmark single-photon sources as it certifies the absence of multi-photon components and quantifies the efficiency simultaneously. Moreover, this probability can be bounded simply from an auto-correlation measurement -- a balanced beam splitter and two photon detectors. Such a bound gives access to various non-classicality witnesses that can be used to certify and quantify Wigner-negativity, in addition to non-Gaussianity and P-negativity of the state produced by the source. We provide tools that can be used in practice to account for an imperfect beam splitter, non-identical and non-unit detection efficiencies, dark counts and other imperfections, to take finite statistical effects into account without assuming that identical states are produced in all rounds, and optionally allow one to remove the detector inefficiencies from the analysis. We demonstrate the use of the proposed benchmark, non-classicality witness and measure using a heralded single-photon source based on spontaneous parametric down-conversion. We report on an average probability that a single photon is produced $\geq 55\%$ and an average measure of the Wigner negativity $\geq 0.004$ with a confidence level of $1-10^{-10}$.
\end{abstract}

\maketitle

\section{Introduction}
Single-photon sources~\cite{Eisaman11,thomas2021race} are key resources for quantum communication~\cite{Sangouard12}, photonic quantum computation~\cite{Kok07} or radiometry~\cite{Chunnilall14,KUCK2021}. Not all single-photon sources are alike and to be scaled up, most applications require efficient sources of true single photons (single-photon Fock/number states). The quality of single-photon sources is usually quantified from an auto-correlation measurement~\cite{Walls08}, that is, by sending the photons to a balanced beam splitter and checking that the ratio between the detected twofold coincidences and the product of singles vanishes, see Fig.~\ref{fig:schemeg2} and 
the discussion in Appendix~\ref{app: autocorrelation}. This ensures that the source produces no more than one photon. The result is, however, insensitive to loss as the efficiency cancels out of the ratio. These two aspects -- the capacity of a source to produce no more than one photon and its efficiency -- are thus considered separately. Both aspects are, however, important and are quantified jointly by the probability that the source actually produces exactly a single photon. Characterizing this probability is a direct and more complete way to benchmark single-photon sources.
\smallskip

\begin{figure}
\centering
\includegraphics[width=0.8\columnwidth]{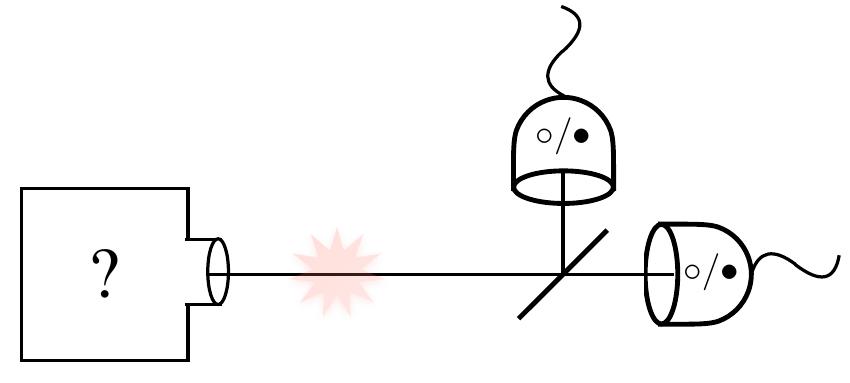}
\caption{Schematic representation of the measurement that is considered to characterize an unknown photon source which supposedly produces single photons. It is realized with a beam splitter and two non-photon-number-resolving detectors, as in a standard auto-correlation measurement. At each round, each detector either clicks $\bullet$ or not $\circ$. By analyzing the frequency of these events, the probability that the source actually produces exactly a single photon can be lower bounded. Furthermore, if a single radiation mode is detected, various forms of non-classicality can be witnessed and quantified.}
\label{fig:schemeg2}
\end{figure}

Interestingly, this probability can be bounded by reconsidering the statistics of detector counts in an auto-correlation measurement. This suggests a systematic way to benchmark the quality and quantify the efficiency of single-photon sources, and to witness and quantify their quantum nature. To motivate benchmarking single-photon sources by the probability that the source actually produces exactly a single photon, we provide a detailed analysis which includes a simple statistical tool to account for finite-size effects without assuming that identical states were produced in all rounds of the experiment. We show how to include imperfections in the measurements apparatus and how to remove the detector efficiencies from the analysis to facilitate the use of the proposed benchmark. An experimental demonstration is presented, illustrating the quality of heralded single-photon sources based on spontaneous parametric down-conversion. 

The auto-correlation measurement is also known to be valuable for witnessing various forms of non-classicality, including the non-positivity of the P-distribution and quantum non-Gaussianity~\cite{Filip112,Predojevic14,Straka2014}. Here we show that for single-mode states or measurements a bound on the single-photon probability can be readily used to quantify the negativity of the Wigner representation~\cite{Walls08}, the strongest form of non-classicality, with respect to a measure proposed  in~\cite{kenfack2004negativity}. We apply this method to verify Wigner-negativity in the reported experiments.

\section{Measurement apparatus}
\label{sec: apparatus}
The measurement apparatus we consider is similar to the one used for the second order auto-correlation measurement. It is a simple measurement consisting of  sending the photonic state to be measured (labeled $\rho$) on to a beam splitter and recording the photon-count correlations between the two outputs, see Fig.~\ref{fig:schemeg2}. We consider that photon detections are made with typical non-photon-number-resolving detectors. In order to draw conclusions from such photon-counts, we introduce a simple quantum model for such a measurement setup. A non-photon-number-resolving detector of efficiency $\eta$ can be modeled with a two element positive operator-valued measure (POVM) $\{E_\bullet, E_\circ\}$  corresponding to click ($\bullet$) and no-click ($\circ$) outcomes. When the measurement acts on a single mode characterized by bosonic operators $a$ and $a^\dag$, the POVM elements take the following form
\be\label{eq: single POVM}
E_\circ = (1-\eta)^{a^\dag a}, \qquad E_\bullet = \id -(1-\eta)^{a^\dag a}.
\ee
This model describes a non-photon-number-resolving detector, for which every incident photon can trigger the detection event with probability $\eta$. The no-click outcome then corresponds to the event where none of the incident photons triggered a click, and occurs with probability $p_\circ = \tr \rho \, (1-\eta)^{a^\dag a}$, see e.g.~\cite{Migdall13}. From now on we assume that the detectors are accurately described by Eq.~\eqref{eq: single POVM} for \textit{some value} of $\eta$. In section \ref{sec: dark counts} we will discuss how to account for small deviation from this model.


When two such detectors are placed after a beam splitter with reflectance $r$ (transmittance $t$), it is straightforward to see that the four possible outcomes are given by the POVM elements
\be\begin{split} \label{eq: POVM elements}
E_{\circ \circ} &= (1-\eta)^{a^\dag a} \\
E_{\bullet \circ} &= (1-\eta\, t)^{a^\dag a}- (1-\eta)^{a^\dag a}\\
E_{\circ \bullet} &= (1-\eta\, r)^{a^\dag a}- (1-\eta)^{a^\dag a}\\
E_{\bullet \bullet} & = \id - E_{\circ \circ} -E_{\bullet \circ} - E_{\circ \bullet},
\end{split}
\ee
where the first (second) label $\circ/\bullet$ refers to the detector after the reflected (transmitted) output of the beam splitter, see Appendix~\ref{sec:povm} for a formal derivation. The events where a fixed detector does not click are modeled by the two POVM elements $E_{\circ\_} = E_{\circ \bullet} + E_{\circ \circ } = (1-\eta\, r)^{a^\dag a}$ and $E_{\_\circ} = (1-\eta \, t)^{a^\dag a}$. The corresponding probabilities are labeled $p_{\circ\_}$ and $p_{\_\circ}$. Note that the case where the two detectors do not have the same efficiency $\eta_{R} \neq \eta_{T}$ can be accounted for by replacing $t$ with $t'=\frac{t\eta_T}{t\eta_T+r\eta_R}$, $r$ with $r'=\frac{r\eta_R}{t\eta_T+r\eta_R}$, and setting $\eta = t\eta_T+r\eta_R$ in Eq.~\eqref{eq: POVM elements}. 

\section{Benchmarking a single-photon source}
\label{sec: benchmark}

For simplicity, through the main part of this section we will be considering the case of single-mode sources. In section \ref{sec:multimode generalzation}, we show that all the presented tools also apply to multi-mode sources.

With the measurement we just described, any single-mode state $\rho$ incident on the beam splitter can be associated with a probability vector 
\begin{equation}
    \bm p = (p_{ \circ\circ},p_{\bullet \circ},p_{\circ \bullet},p_{\bullet\bullet})
\end{equation} 
governing the occurrence of clicks. Our goal is to construct an estimator $\hat{P}_1({\bm p})$ that relates this vector $\bm p$ to the photon number statistics of the state and in particular to the weight of the single-photon component $P_1 = \bra{1} \rho \ket{1}$. Directly lower bounding $P_1$ is a natural way to benchmark a single-photon source. In particular, it sets a bound on the trace distance between the state $\rho$ prepared by the source and an ideal single photon $\ket{1}$
\be
\frac{1}{2}\left\| \rho -\ketbra{1}{1}\,\right\|_1 \leq 1- P_1,
\ee
which can be readily used to bound the errors for various applications of single-photon sources.

\subsection{Measurement calibration independent benchmark}

We proceed step by step, considering first an ideal measurement apparatus, then considering two identical non-unit detector efficiencies and finally focusing on the most imperfect measurement, where we consider an unbalanced beam splitter and two detectors having different efficiencies. At this point we only assume that the measurement is described by the POVM in Eq.~\eqref{eq: POVM elements} for \emph{some} parameters $\eta,t$ and $r$, that do not need to be known. The benchmark that we will derive in this section is thus independent of the calibration of the measurement apparatus. Finally, we show that by adding a small correction term our benchmark can be applied in situations where the POVM in Eq.~\eqref{eq: POVM elements} is only an approximate description of the measurement apparatus.

\subsubsection{Ideal measurement apparatus}
\label{sec: benchmark ideal}
For clarity, we first assume that the source produces an identical state $\rho$ at each round, the detectors have unit detection efficiency $\eta=1$ and the beam splitter is balanced $t=r=1/2$. 
In this ideal case, the probabilities $p_{\bullet \circ}$, $p_{\circ \bullet}$ are equal, and $\bm p$ is described by two independent real parameters.
For convenience, we introduce $p_\bullet=p_{\bullet \circ}+ p_{\circ \bullet}$ the probability to get exactly one click. The probabilities that a given detector does not click $p_{\circ\_}=p_{\circ \circ}+ p_{\circ \bullet}$ and $p_{\_\circ}=p_{\circ \circ}+ p_{\bullet \circ}$ are equal in the ideal case and in particular  $p_{\circ\_}= p_{\_\circ} = p_{\circ\circ}+ \frac{1}{2}p_\bullet$. This means that the probabilities of outcomes of the measurement of interest can be fully captured by $(p_{\circ\_}, p_{\circ\circ})$.
Let 
\be
P_n =\bra{n}\rho \ket{n}
\ee
be the weight of the $n$-photon Fock state component of the measured state. For $\eta=1$, the no-click events can only come from the vacuum state $E_{\circ\circ} =\prjct{0}$, hence the probabilities $(p_{\circ\_}, p_{\circ\circ})$ can be  linked to the photon number distribution $P_n$. From Eqs.~\eqref{eq: POVM elements}, one  gets $p_{\circ\circ} = P_0$ and $p_{\circ\_} = \sum_n P_n \frac{1}{2^n}$.
\smallskip

The question we ask now is very simple -- what are the values $(p_{\circ\_}, p_{\circ \circ})$ that are obtainable for states $\rho$ satisfying $P_1\leq P$, for some parameter $P \in [0,1]$?

\begin{figure}[t]
\centering
\includegraphics[width = \columnwidth]{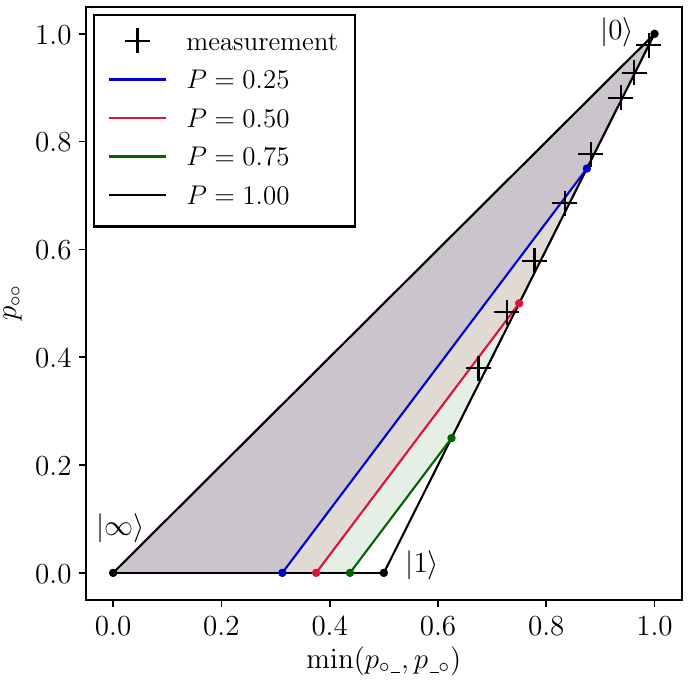}
\caption{\label{fig:measurement} 
Representation of the polytopes $\bm Q_{P}$ (defined after Eq.~\eqref{eq: p_o hat}) containing all possible values $(\min(p_{\circ\_}, p_{\_\circ}),p_{\circ \circ})$ associated to states  $\varrho$ with $\bra{1}\varrho\ket{1}\leq P$. There are four polytopes $\bm Q_{P}$ for the values $P\in\{0.25,0.5,0.75, 1\}$. The physically possible region $\bm Q_{P=1}$ is given by the black triangle. The regions for $P=0.25,0.5,0.75$ are given by the part of the black triangle above the corresponding colored line. The black crosses are measurements of $(\min(p_{\circ\_}, p_{\_\circ}),p_{\circ \circ})$ for a heralded single photon undergoing different added attenuation corresponding to transmission efficiencies of $\eta_\text{att}\in \{1.0, 0.83, 0.68, 0.51, 0.36, 0.19, 0.12, 0.034\}$.}
\end{figure}

 First, we note that $p_{\circ\_}\geq p_{\circ \circ}$ holds by definition. Furthermore, the points $(1,1)$ and $(0,0)$  are attained by the vacuum and the state with infinitely many photons, respectively. Thus, the line  $p_{\circ\_} = p_{\circ \circ}$ is also attainable by mixtures of aforementioned states. Then,
 we look for the maximum value of $p_{\circ\_} =\sum_n P_n \frac{1}{2^n} $ for a fixed $p_{\circ\circ}$. We have to solve
 \be\begin{split}
p_{\circ\_}^{\uparrow}(p_{\circ\circ},P) =\max_{\rho}   &\sum_n P_n \frac{1}{2^n}\\
 \t{such\, that} \quad & P_1\leq P\\
 & P_0 =p_{\circ\circ}.
 \end{split}
 \ee
As $(1/2)^n$ is decreasing with $n$, the maximum is attained by saturating the values of $P_n$ starting with $P_0$. Hence, it equals
\be\label{eq: p_o hat}
p_{\circ\_}\leq p_{\circ\_}^{\uparrow}(p_{\circ\circ},P) =
\begin{cases} \frac{1+p_{\circ \circ}}{2} & 1-p_{\circ \circ}\leq P\\
\frac{1+P+3 p_{\circ \circ}}{4} & 1-p_{\circ \circ}>P
\end{cases}\ee
The set of possible values $(p_{\circ\_}, p_{\circ \circ})$ is thus included in a convex polytope with four vertices 
$\bm Q_P = \t{Polytope} \{(0,0),\left(\frac{1+P}{4},0\right), \left(\frac{2-P}{2},1-P \right), (1,1) \}$, sketched in Fig.~\ref{fig:measurement}. The only nontrivial facet of this polytope is the edge connecting $\left(\frac{1+P}{4},0\right)$ and $\left(\frac{2-P}{2},1-P \right)$ which is associated to the inequality $4 p_{\circ\_}- 3 p_{\circ \circ} -1 \leq P$, and is given by the colored lines in Fig.~\ref{fig:measurement}. Thus, without loss of generality, the condition $\bra{1} \rho \ket{1}\leq P$ implies that the elements of $\bm p$ satisfy the linear constraint  
\be
\label{eq:benchamrk}
\hat{P}_1^T({\bm p}) = 
4 p_{\circ\_}- 3 p_{\circ \circ} -1 \leq P.
\ee
Conversely, by measuring the pair $(p_{\circ\_},p_{\circ \circ} )$ and by computing the resulting value of $\hat {P}_1^T$, we can guarantee that for any value of $P$ such that $\hat {P}_1^T >P$ ,  $\bra{1}\rho \ket{1} > P$ holds, that is we get a lower bound on the probability that the  source to be benchmarked produces exactly a single photon.

\subsubsection{Identical non-unit efficiency detectors}
To move away from the ideal case, we still consider a perfectly balanced beam splitter and focus on a situation where non-unit efficiency detectors are used. We consider the case where the detector efficiency $\eta$ is unknown. In the measurement setup shown in Fig.~\ref{fig:schemeg2}, non-unit efficiency detectors can be modeled by taking ideal detectors and placing a beam splitter with transmission $\eta$ before the balanced beam splitter. As a consequence, observing a violation of Ineq.~\eqref{eq:benchamrk} proves that the state produced by the single-photon source  \emph{and undergoing losses} satisfies $\bra{1}\varrho \ket{1}\geq \hat{P}_1^{T}({\bm p})$. This provides a valid  benchmark even though the intrinsic quality of the source is estimated with a lossy measurement apparatus. It is interesting to note that for any state $\rho$ with $P_1 \geq 2/3,$ the probability of the single-photon weight $P_1$ can only decrease with loss, see Appendix~\ref{sec:loss}. Therefore, showing that $P_1\geq 2/3$ with lossy detectors implies that the original state also satisfies $P_1\geq\frac{2}{3}$. 
This is not the case when $P_1<2/3$, i.e. for specific states the single-photon weight $P_1$ can be increased by loss (an intuitive example is the two-photon Fock state).

\subsubsection{Unbalanced beam splitter and different non-unit efficiency detectors}
We now relax the assumptions that the beam splitter is balanced and the detector efficiencies are the same, i.e. we consider the case with a measurement performed with a beam splitter having an unknown transmission $t$ and reflection $r=1-t$ and two detectors having different efficiencies labeled $\eta_R$ and $\eta_T$. In this case, the observed statistics would be equivalently obtained with a beam splitter having a transmission coefficient $t'=t \eta_T /(t \eta_T + r \eta_R)$ and two detectors with the same efficiency $\eta=t\eta_T+r\eta_R$, as already mentioned below Eq.~\eqref{eq: POVM elements}. This means that the measurement can be modeled with a first unbalanced beam splitter with transmission coefficient $\eta$ corresponding to loss on the state to be characterized, an unbalanced beam splitter with transmission $t'$ and two detectors with unit detection efficiency. In this case, the relation between $p_{\circ\circ}$ and $P_0$ is unchanged, i.e. $p_{\circ\circ} = P_0$. The probabilities $p_{\circ\_}= p_{\circ \bullet} + p_{\circ\circ}$ and $p_{\_\circ}= p_{\bullet \circ } + p_{\circ\circ}$
 are however no longer the same. They are now given by $p_{\circ\_} = \sum_n P_n (1-r')^{n}$ and $p_{\_\circ} = \sum_n P_n (1-t')^{n}$. Hence, the quantity $\sum_n P_n \frac{1}{2^n}$ is no longer directly related to the probability $\bm p$. Nevertheless, it can be bounded from observable quantities, as $\sum_n P_n \frac{1}{2^n} \geq \min(p_{\circ\_}, p_{\_\circ})$.

We introduce $\hat{P}_1^R({\bm p})$ which is defined analogously to $\hat{P}_1^T({\bm p})$ by $\hat{P}_1^R({\bm p}) = 4 p_{\_\circ}- 3 p_{\circ \circ} -1$. Using  the definition of $p_{\circ\_}$ and $p_{\_\circ}$, we rewrite them in terms of probabilities of disjoint events as
\begin{equation}
    \label{eq: general witness}
\begin{split}
    &\hat{P}_1^T(\bm p)=4 p_{\circ \bullet}  + p_{\circ\circ}-1 \\
    &\hat{P}_1^R(\bm p)=4 p_{\bullet \circ}  + p_{\circ\circ}-1.
\end{split}
\end{equation}
With this notation in hand, we conclude that the quantity
\be
\label{benchmark}
 \hat{P}_1({\bm p})= \min\{ \hat{P}_1^T({\bm p}), \hat{P}_1^R({\bm p})\}
\ee 
is a benchmark for single-photon sources, without assumptions on the detector efficiencies and on the fact that the beam splitter is balanced.
This means that from the outcome probabilities $\bm p$ of a usual auto-correlation measurement, we can compute $\hat{P}_1^T({\bm p})$ and $\hat{P}_1^R({\bm p})$, deduce their minimum $\hat{P}_1({\bm p})$ and guarantee the tested source produces states with the weight of the single-photon component satisfying $P_1=\bra{1}\rho \ket{1} \geq \hat{P}_1({\bm p})$.

\subsubsection{Dark counts and other imperfections}

\label{sec: dark counts}
Finally let us briefly consider general passive detectors described by a POVM 
\be
\tilde E_\circ = \sum_{n\geq 0} e_\circ(n) \ketbra{n}
\quad \text{and} \quad \tilde E_\bullet =  \sum_{n\geq 0}  e_\bullet(n) \ketbra{n},
\ee
where $e_\circ(n)+e_\bullet(n)=1$. Consider combining two such detectors in the $g^{(2)}$ setup, with the POVM $\{\tilde E_{\circ/\bullet}\}$ performed on the transmitted mode and $\{\tilde E'_{\circ/\bullet}\}$ on the reflected one. The resulting POVM elements for $a,b =  \circ, \bullet$ read
\be\label{eq: general g2 POVM}\begin{split}
\tilde E_{a b} = \sum_{n\geq 0} \sum_{k=0}^n &\text{P}_\text{BS}(k|n) \\
&e_a(k) e_b'(n-k) \ketbra{k,n-k}.
\end{split}
\ee
with $\text{P}_\text{BS}(k|n) = \binom{n}{k} t^k r^{n-k}$. 

Now let us assume these detectors are not too different from the textbook single-photon detector model of Eq.~\eqref{eq: single POVM}. Concretely we assume that for some $\eta, t$ and $r$
\be\label{eq: ound POVM}
\left|\tr (\tilde E_{ab}- E_{ab})\rho \right| \leq \Delta,
\ee
where $\rho$ can be any state susceptible to be prepared in the experiment. In particular, consider the case where both detectors satisfy 
\be\begin{split}
|e_\circ(n)-(1-\eta)^n|&\leq \delta\\
|e'_\circ(n)-(1-\eta')^n|&\leq \delta'
\end{split}
\ee 
for some $\eta,\eta'$, and all $n$ on which the input state is supported. For $a =b =\circ$ we then find
$|e_\circ(k) e_\circ'(n-k)- (1-\eta)^k(1-\eta')^{n-k}|\leq \delta+\delta'+\delta\delta'$ in Eq.~\eqref{eq: general g2 POVM}.  Similar inequalities hold for the other  possible outcomes $\circ\bullet$ ,$\bullet\circ$ and $\bullet\bullet$, and lead to the bound of Eq.~\eqref{eq: ound POVM} with
\be\label{eq: DELTA}
\Delta =\delta+\delta'+\delta\delta'.
\ee

It follows that the probabilities $\tilde {\bm p}=(\tilde p_{\circ \circ}, \tilde p_{\circ \bullet}, \tilde p_{\bullet \circ}, \tilde p_{\bullet \bullet})$ observed with these detectors described by $\{\tilde E_{\circ/\bullet}\}$ and $\{\tilde E'_{\circ/\bullet}\}$, are close to the probabilities $\bm p=( p_{\circ \circ},  p_{\circ \bullet},  p_{\bullet \circ}, p_{\bullet \bullet})$ that would have been observed with textbook detectors, concretely
\be
|\tilde p_{ab} - p_{ab}|\leq \Delta.
\ee
With the help of Eq.~\eqref{eq: general witness} it is then straightforward to lower bound the value of the benchmark
\be\label{eq: general benchmark}
 \hat{P}_1({\bm p}) \geq \min\{ \hat{P}_1^T({\tilde{\bm p}}), \hat{P}_1^R({\tilde{\bm p}})\} - 5 \Delta.
\ee
given the coincidence probabilities $\tilde{\bm p}$ observed with any detectors abiding to Eq.~\eqref{eq: ound POVM}.

Here, it is worth noting that the POVM model of Eq.~\eqref{eq: single POVM} does not account for dark counts. These can be modeled by setting a nonzero click probability for the vacuum $e_\bullet(0)=p_\text{dc}$, implying $|e_\circ(n)-(1-\eta)^n|\leq \delta =p_\text{dc}$. Then accordingly to Eqs.~(\ref{eq: DELTA},\ref{eq: general benchmark}), to account for the effect of dark counts one can use the bound
\be\label{eq: corr dark counts}
\hat{P}_1({\bm p}) \geq \min\{ \hat{P}_1^T({\tilde{\bm p}}), \hat{P}_1^R({\tilde{\bm p}})\} - 10 \, p_\text{dc},
\ee
where we only took the leading order correction in $p_\text{dc}$, which is typically fairly small. 

\subsection{Measurement calibration dependent benchmark}

\label{sec: calibration deendent benchmark}

The benchmark we proposed relies on no assumption on the characteristics of the beam splitter or the two detectors used in the auto-correlation measurement. This prevents a miscalibration of the measurement apparatus that could result in an overestimation of the quality of the single-photon source. Nevertheless, as one would expect, the bound on the quality of the tested source (see Eq.~\eqref{benchmark}) is reduced when using  an unbalanced beam splitter or inefficient detectors. 
We now discuss a way to characterize the intrinsic quality of the source by making additional assumptions on the measurement apparatus. The basic idea is to exploit estimations of different losses in photonic experiments.
In particular, we assume that the detector efficiencies $\eta_{R(T)}$ and the beam splitter reflectivity $r$ are bounded, that is  $\eta_{R(T)} \leq \hat{\eta}_{R(T)}$ and $r\in [1-\hat t,\hat r]$. We show in Appendix~\ref{sec:param} that the condition $\bra{1} \rho \ket{1}\leq P$ implies that the two following inequalities hold 
\be
\label{parambound}
\begin{split}
\hat{P}_1^{T\ast}(\bm p) &= C_1(\hat t,\hat \eta_T) p_{\circ \bullet} -C_2(\hat t,\hat{\eta}_T,\hat{\eta}_R) p_{\bullet \bullet} \leq P,\\
\hat{P}_1^{R\ast}(\bm p) &= C_1(\hat r,\hat \eta_R) p_{\bullet \circ} - C_2(\hat r,\hat{\eta}_R,\hat{\eta}_T) p_{\bullet \bullet} \leq P,
\end{split}
\ee
where the coefficients $C_1$ and $C_2$, defined as
\be\label{eq: C1 C2}
\begin{split}
C_1(x,\eta) &=\frac{1}{x\, \eta}\\
C_2(x,\eta_1,\eta_2) & = \frac{1}{x\, \eta_1}\left(\frac{2-x\, \eta_1}{2(1-x)\eta_2}-1\right)\\
\end{split}
\ee
have been optimized such that $\hat{P}_1^{T\ast}(\bm p) $ and $\hat{P}_1^{R\ast}(\bm p) $ give the tightest bound on $P$.
Finally, one can choose the best among the two bounds, giving rise to the benchmark
\be
\label{benchmark2}
 \hat{P}_1^\ast({\bm p})= \max\{ \hat{P}_1^{T\ast}(\bm p),\hat{P}_1^{R\ast}(\bm p)\} \leq P_1
\ee 
for the single-photon probability,
which takes advantage of the additional experimental knowledge.

Again we can account for  deviations of the POVMs $\{\tilde E_{\circ/\bullet}\}$ and $\{\tilde E'_{\circ/\bullet}\}$ describing the detectors form the textbook model of Eq.~\eqref{eq: single POVM}, as quantified by Eq.~\eqref{eq: DELTA}. Repeating the analysis of Sec.~\ref{sec: dark counts}, we obtain the bounds 
\be\label{eq: dark count param}\begin{split}
\hat{P}_1^{T\ast}(\bm p) &\geq \hat{P}_1^{T\ast}(\tilde{\bm p}) - (C_1(\hat t,\hat \eta_T) + |C_2(\hat t,\hat{\eta}_T,\hat{\eta}_R)|)\Delta \\
\hat{P}_1^{R\ast}(\bm p) &\geq \hat{P}_1^{R\ast}(\tilde{\bm p}) -
(C_1(\hat r,\hat \eta_R)+ |C_2(\hat r,\hat{\eta}_R,\hat{\eta}_T)| )\Delta.
\end{split}
\ee
To account for dark counts one may use $\Delta = 2 p_\text{dc}$. Recall that here $\tilde {\bm p} =(\tilde p_{\circ \circ}, \tilde p_{\circ \bullet}, \tilde p_{\bullet \circ}, \tilde p_{\bullet \bullet})$ is the vector of probabilities observed with the real detectors, while $\bm p$ gives the probabilities that would have been observed with the textbook model and that are used in the benchmark.

\subsection{Multi-mode sources}
 \label{sec:multimode generalzation}

We have so far considered sources emitting light in a single mode, or equivalently that the emitted light is filtered in all the auxiliary degrees of freedom so that a single mode of light is detected.
We will now briefly consider the situation where the detected state is multi-mode, each mode being associated to an annihilation operator $a_k$ satisfying $[a_k,a_\ell^\dag] = \delta_{k\ell}$. 
The no-click and click events for a multi-mode input are associated to the POVM elements $E_\circ = \bigotimes_k (1-\eta)^{a_k^\dag a_k}$ and $E_\bullet =\id-E_\circ$ -- a detector does not click only if none of the modes triggers a click. To a multi-mode state $\rho$ one associates the distribution $P_n$ of the \emph{total photon number} operator $\hat n=\sum_k a^\dag_k a_k$. Assuming that the beam splitter acts identically on all modes, the measurement apparatus is only sensitive to the total number of photons $\hat n$. That is, the POVM elements $\{E_{\circ\circ}, E_{\circ\bullet}, E_{\bullet\circ},E_{\bullet \bullet} \}$ are given by Eq.~\eqref{eq: POVM elements} albeit with $a^\dag a$ replaced by $\hat n$, see Appendix~\ref{sec:povm}. We thus conclude that the quantities $\hat P_1(\bm p)$ and $ \hat{P}_1^*(\bm p)$ that have been derived, can be readily used to benchmark the probability that a multi-mode source emits a single photon 
\be
P_1 = \tr \left(\varrho \sum_k a_k^\dag \ketbra{0}{0} a_k\right).
\ee

It is worth noting that a high $P_1$ for a multi-mode source does not guarantee that the single-photon probability is high in any of the individual modes. As a practical example, consider the multi-mode single-photon state 
\be\label{eq: single photon multimode}
 \varrho = \frac{1}{N}\sum_{j=1}^N a_j^\dag \ketbra{0}{0} a_j.
\ee
 This state has exactly one photon in total $\hat n =1$. Nevertheless, it is not a good \emph{single-mode} single-photon state. The probability to find exactly one photon in any single mode is only $1/N$, therefore two such states would only exhibit a limited two-photon interference (bunching). In other words $\varrho$ is not a pure single-photon state $\ketbra{1}{1}_1$, the degree of freedom that distinguishes between the different modes is in a highly mixed states. Depending on the application, the mode-purity (or single-mode character) of the source may be either irrelevant (e.g. for some quantum random number generators) if the interference between different sources plays no role, or crucial (e.g. quantum repeaters, boson sampling or photonic quantum computation) if it is at the heart of the task.

 In Appendix~\ref{app:multimode}, we show that  one can guarantee that the source produces a single-mode state with high $P_1$ if it is reasonable to assume that the multi-mode state is a product $\varrho = \bigotimes_k \rho_k$. In general, however, the auto-correlation measurement is intrinsically insensitive to the multi-mode characteristics of the source, as illustrated by the example of Eq.~\eqref{eq: single photon multimode}. In principle, it is possible to extract a single-mode state from a multi-mode source by filtering the auxiliary degrees of freedom, e.g. using a single-mode fiber for spatial degrees of freedom and a spectral filter for frequency domain. In any case, the mode-purity of the detected state has to be verified with a different set of measurements, e.g. via Hong-Ou-Mandel interference~\cite{cassemiro2010} or based on a physical model of the source. For example, in the case of a heralded single-photon source, the spectral purity can be determined by measuring the signal-idler joint spectral intensity~\cite{Zielnicki2018}, as will be explained in Sec.~\ref{sec: exp wigner}. 
 
 A systematic analysis of the characterization of the mode-purity of a source is beyond the scope of this paper. Nevertheless, it is worth mentioning  that with the additional information on the multi-mode characteristics of the source and the knowledge of the total single-photon probability $P_1$ it is possible to bound the probability to find a single photon in the mode of interest $P_1^{[1]}$ (denoted mode 1). In particular, given a bound on the probability that all the modes but one are empty  $\tr \varrho (\id_1\otimes\ketbra{0}_{2}\otimes \dots\otimes \ketbra{0}_N) \geq 1-\varepsilon$, one can conclude that
\be
\label{eq: multi-mode P}
P_1^{[1]} = \tr \prjct{1}_1 \rho_k\geq P_1 -\varepsilon,
\ee
since the maximal contribution to $P_1$ from the other modes is $\varepsilon$. Here, $\rho_1 = \tr_{2,\dots,N} \varrho$ is the marginal state of the first mode, which can in principle be filtered from the source. A bound on $P_1^{[1]}$ can then be used to, e.g., quantify the Wigner-negativity of the state $\rho_1$, see Sec.~\ref{sec: wigner}.

\section{Finite statistics}
\label{sec: finite statistics}

In this section, we analyze finite size effects for the benchmark.

We sketch an analysis to account for finite statistics in any experiment aiming to evaluate $\hat{P}_1({\bm p})$, the benchmark for single-photon sources derived in Sec.~\ref{sec: benchmark}. 
For a measurement round described by $\bm p$ we associate a random variable $X_T$ that takes different real values depending on the measurement result
\be
X_T = \begin{cases}
3 & (\circ \bullet) \\
0 & (\circ\circ) \\
-1 & (\bullet\circ) \,\, \textrm{or} \,\, (\bullet \bullet)
 \end{cases}
\ee
This random variable satisfies $\mathds{E}(X_T) = \hat{P}_1^T(\bm p)$ in Eq.~\eqref{eq: general witness} (here and further $\mathds{E}$ denotes the expected value of a random variable). Analogously, we define $X_R$ by exchanging the role of the two detectors, such that it satisfies $\mathds{E}(X_R) = \hat{P}_1^R(\bm p)$.

In general, the source may prepare a different state $\rho^{(i)}$ at each round, corresponding to different probabilities $\bm p^{(i)}$ of the measurement outcomes. This means that in each round, we sample different random variables $X_T^{(i)}$ and $X_R^{(i)}$, which are independent between rounds given the sequence of states $\rho^{(1)},\dots,\rho^{(n)}$ produced in the experiment. In this case, a reasonable figure of merit is the average quality of the state prepared by the source $\bar P_1 = \frac{1}{n} \sum_{i=1}^n P_1^{(i)}$ where $P_1^{(i)}$ is the probability of the single-photon component of the state $\rho^{(i)}$. Because the transmission and reflection coefficients of the beam splitter can be considered to be constant, either $\mathds{E}(X_T^{(i)}) \geq \mathds{E}(X_R^{(i)})$ or $\mathds{E}(X_T^{(i)}) \leq \mathds{E}(X_R^{(i)})$ holds for all $i$. This means that the average single-photon weight fulfills
\be
 \bar P_1 \geq \min\{ \mathds{E}(\bar X_T), \mathds{E}(\bar X_R) \},
\ee
where $\bar X_{T(R)} = \frac{1}{n} \sum_{i=1}^n X_{T(R)}^{(i)}$. Finally, we use the Hoeffding 1963 theorem~\cite{Hoeffding1963} to show  that  
\be
\label{eq:q_alpha}
\hat q_\alpha  =    \min \{ \bar X_T, \bar X_R\} - \sqrt{\frac{16 \log(1/\alpha)}{2 n}}
\ee
is a one-sided confidence interval
on $\bar P_1$ with confidence $\alpha$ (see Appendix~\ref{sec:stat}). Precisely, with probability $1-\alpha$ the observed value of $\hat q_\alpha$ lower bounds $\bar P_1$. It might be convenient to note that the quantity $\min \{ \bar X_T, \bar X_R\}$ of this confidence interval can be computed using
\be
\min\{\bar X_T,\bar X_R\} = \frac{4 \min \{n_{\circ \bullet},n_{\circ \bullet} \} - n_{\circ \bullet}- n_{ \bullet \circ} - n_{\bullet \bullet}}{n}
\ee
with e.g. $n_{\bullet \bullet}$ counting the number of outcomes $\bullet \bullet$. \\

Analogously, one can derive a one-sided confidence intervals associated to the calibration-dependent benchmark derived in Sec.~\ref{sec: calibration deendent benchmark}. With very similar arguments one can show that both
 \be\begin{split}
\hat q_\alpha^{(T)\ast} &= \frac{ C_1(\hat t,\hat \eta_T) n_{ \circ\bullet }- C_2(\hat t,\hat{\eta}_T,\hat{\eta}_R) n_{\bullet \bullet}}{n}\\ 
& \quad - (C_1(\hat t,\hat \eta_T)+C_2(\hat t,\hat{\eta}_T,\hat{\eta}_R))\sqrt{\frac{\log(1/\alpha)}{2 n}},\\
\hat q_\alpha^{(R)\ast} &= \frac{ C_1(\hat r,\hat \eta_R) n_{\bullet \circ}-C_2(\hat r,\hat{\eta}_R,\hat{\eta}_T) n_{\bullet \bullet}}{n}\\ 
& \quad - (C_1(\hat r,\hat \eta_R)+C_2(\hat r,\hat{\eta}_R,\hat{\eta}_T))\sqrt{\frac{\log(1/\alpha)}{2 n}},\\
 \end{split}
\ee
where the functions $C_1$ and $C_2$ defined in Eq.~\eqref{eq: C1 C2} are confidence intervals for $\bar P_1$. Since they are derived from the same data, for a statistically meaningful statement one has to chose parameter $\alpha$ before computing the confidence interval. On the other hand, it is straightforward to see that
 \be\label{eq: q*}
\hat{q}_{2\alpha}^{\ast} =\max \{q_\alpha^{(T)\ast},q_\alpha^{(R)\ast} \} 
\ee
is also a confidence interval at confidence level $1-2\alpha$. See App.~\ref{sec:stat} for a detailed derivation.

\section{Relation to the non-classicality of the source}
\label{sec: wigner}

The data obtained from auto-correlation type measurements are known to be valuable for witnessing and quantifying various forms of non-classicality, including the non-positivity of the P-function and quantum non-Gaussianity~\cite{Filip112,Predojevic14,Straka2014}. 
We now show that the knowledge of $P_1$ in a single bosonic mode (as e.g. provided by our benchmarks $\hat P_1(\bm p)$ and $\hat{P}_1^*(\bm p)$) can reveal Wigner-negativity~\cite{Wigner32}, arguably the strongest form of non-classically for a bosonic mode. In particular, Wigner-negativity implies the non-positivity of the P-function~\cite{vogel2006quantum}. 
Similarly, it implies that the corresponding state is non-Gaussian, as a Gaussian state has a Gaussian (and thus positive) Wigner function\footnote{In addition, Hudson's theorem \cite{HUDSON1974} tells us that any pure state with a positive Wigner function is Gaussian.}.
Thus, demonstrating Wigner-negativity for a light source brings  evidence of its strong quantum nature. Note that it has been shown recently that witnesses of Wigner-negativity can be derived systematically using a hierarchy of semidefinite programs~\cite{Chabaud21}. Our contribution is more specific and aims at witnessing Wigner-negativity simply and directly from $\hat P_1(\bm p)$ or $\hat{P}_1^*(\bm p)$ observed on a single mode.

\smallskip
\subsection{Wigner-negativity witness}
\label{sec: wigenr witness}

The Wigner function is a representation of a single-mode state $\rho$ in terms of the following quasi-probability distribution~\cite{Royer77}
\be
\label{wigner}
W_{\rho}(\beta)=\frac{2}{\pi}\text{Tr} (\t{D}_\beta(-1)^{a^{\dagger}a}\t{D}_\beta^\dag\, \rho),
\ee
with $\int \dd \beta^2 W_{\rho}(\beta)=1$. 
Here, $\t{D}_\beta=e^{a^\dag \beta - a \beta^*}$ is the displacement operator with a complex amplitude $\beta$. Applying Eq. \eqref{wigner} to a Fock state gives~\cite{vogel2006quantum}
\be
\label{eq: wigner Fock}
W_{\ketbra n}(\beta)=\frac{2(-1)^n}{\pi} e^{-2|\beta|^2} L_n\left(4|\beta|^2\right)
\ee
where $L_n$ is the Laguerre polynomial. Note that the following bound on the Laguerre polynomials $e^{-x/2}|L_n(x)|\leq 1$, see e.g. Eq. (18.14.8) in \cite{NIST}, leads to a bound on the Wigner function of Fock states $|W_{\ketbra n}(\beta)|\leq \frac{2}{\pi}$. Note also that $L_1(x)=1-x$.

With the help of Eq.~\eqref{eq: wigner Fock}, the upper bound on the Wigner function of Fock states and the definition of the Laguerre polynomial $L_1(x)$, it is easy to see that the Wigner function of any mixture of Fock states $\rho = \sum p_n \prjct{n}$ satisfies~\footnote{For a general state $\varrho =\sum_{nm} c_{nm}\ketbra{n}{m}$ with Wigner function $W_\varrho(\beta)$, one can always define the corresponding Fock state mixture $\rho=\sum_n P_n \ketbra{n}$ with $p_n =c_{nn}$. Its Wigner function reads $W_\rho(\beta)= W_\rho(|\beta|)=\int \dd\varphi W_\varrho(|\beta|e^{\ii \varphi})= \mean{W_\varrho(|\beta|e^{\ii \varphi})}_\varphi$. The two functions coincide at the origin $W_\rho(0)=W_\varrho(0)$. Furthermore, $W_\rho$ can only be non-positive if $W_\varrho$ is non-positive, and $N_{W_\rho}\leq N_{W_\varrho}$ (introduced at the end of the section) follows from $|\mean{W(|\beta|e^{\ii \varphi})}_\varphi| \leq \mean{|W(|\beta|e^{\ii \varphi})|}_\varphi$ .}
\be\label{eq: W bound}
\begin{split}
W_\rho (\beta) &= P_1 W_{\ketbra 1}(\beta) + \sum_{n\neq 1} P_n W_{\ketbra n}(\beta)\\
    &\leq  \frac{2}{\pi}\left (- P_1 (1- 4|\beta|^2) e^{-2|\beta|^2} + (1-P_1) \right).
\end{split}
\ee
Focusing on the origin $\beta=0$, we get $ W_\rho (0)\leq 2 \frac{1-2P_1}{\pi}$ which is negative if $P_1$ is larger than $\frac{1}{2}.$ Hence, if one concludes from the measurement of $\bm p$ that $\hat{P}_1({\bm p})>\frac{1}{2}$, one can conclude that the measured state is Wigner-negative (recall that we assumed that the state $\rho$ is single-mode). 

\subsection{Wigner-negativity measure}

A natural way to quantify the negativity of the Wigner representation of a given state $\rho$ is to measure the total quasi-probability for which the function $W_\rho(\beta)$ takes negative values~\cite{kenfack2004negativity}, i.e.
\be
N_W(\rho)= \int \dd\beta^2\frac{|W_\rho(\beta)| -W_\rho(\beta)}{2},
\ee
which is manifestly zero for states with a positive Wigner function.
In the Appendix~\ref{sec:wigneg} we show that $N_W(\rho)$ is non-increasing under Gaussian operations, which justifies its use as a measure of Wigner-negativity. 
Note that with the help of Ineq.~\eqref{eq: W bound}, we show that $N_W(\rho)$ satisfies
\begin{equation}\label{eq: wigner measure}\begin{split}
    N_W(\rho) & \geq F(P_1)=
    \begin{cases}
    \frac{3 (1-P_1) \left(4 w^2+3\right)}{8 w}+\!P_1-\!2 & \!P_1>\frac{1}{2}\\
    0 & \!P_1\leq\frac{1}{2}
    \end{cases}\\
&\text{with} \quad w = w_0\left( \frac{\sqrt{e}}{2}\frac{1-P_1}{P_1}\right),
\end{split}
\end{equation}
 where $w_0$ is the principal branch of the Lambert W function. The function $F(P_1)$ is non-decreasing. Hence, from the measurement of $\bm p,$ we get a lower bound $\hat P_1(\bm p)$ on $P_1$ that can be used to lower bound $N_W(\rho)$ using $F(\hat P_1(\bm p))$. The bound~\eqref{eq: wigner measure} is tight by construction in the ideal case $N_W(\ket{1}) = F(1) =\frac{9}{4\sqrt{e}}-1 \approx 0.36$.

 By computing $F''(P_1)\geq 0$ we show that the  function $F(P_1)$ in Eq.~\eqref{eq: wigner measure} is convex. This property will be used in the following section, where we discuss the finite statistics effects.

\subsection{p-value to witness Wigner-negativity}

We now present the extension of finite statistics analysis presented in Sec.~\ref{sec: finite statistics} to the task of Wigner-negativity detection and quantification. 

First, let us now consider the witness of Wigner-negativity discussed in Sec.~\ref{sec: wigenr witness}, that is 
$W_\rho(0)\geq 0 \implies \hat{P}_1(\bm p) \leq 1/2$, 
and quantify the statistical significance of its contrapositive  given the measurement data. This can be done by computing the p-value
associated to the hypothesis that the Wigner function of the state is positive. As before, we consider the general case where a different state $\rho^{(i)}$ may be prepared at each run. Nevertheless, at each round the bound $W_{\rho^{(i)}}(0)\leq \frac{2}{\pi}(1-2 P_1^{(i)})$ holds. Therefore, for the sequence of states prepared in the experiment, the average Wigner function at the origin satisfies
\be 
\bar{W}(0)=\frac{1}{n}\sum_{i=1}^n W_{\rho^{(i)}}(0) \leq \frac{2}{\pi}(1-2 \bar {P}_1),
\ee
and is negative if $\bar {P}_1 > \frac{1}{2}$. Given some value of $\min\{\bar X_T, \bar X_R\}$  recorded after $n$ measurement rounds, we show in Appendix~\ref{sec:stat} that for any collection of $n$ states with $\bar W(0)\geq 0$, 
the probability that the results are equal or exceed the observed value of $\min\{\bar X_T, \bar X_R\}$ is given by
\be
\label{pvalue}
\text{p-value} \leq \exp\left(-\frac{2 n \left(\min\{\bar X_T, \bar X_R\}-\frac{1}{2}\right)^2}{16}\right),
\ee
for $\min\{\bar X_T, \bar X_R\} > \frac{1}{2}$. 
In other words, given the observed value of $\min\{\bar X_T, \bar X_R\}$, the probability that it is coming from states that are Wigner-positive on average is bounded by the right-hand side of Ineq.~\eqref{pvalue}. In App.~\ref{sec:stat} one finds a bound on the p-value for the calibration-dependent setting. 

\subsection{Confidence interval on the measure of Wigner-negativity}

Finally, the convexity of the function $F(P_1)$ in Eq.~\eqref{eq: wigner measure} implies that the average Wigner-negativity satisfies
$\bar N_W = \frac{1}{n} \sum_i N_W(\rho^{(i)})\geq \frac{1}{n} \sum_{i}F(P_1^{(i)}) \geq F(\bar P_1)$. Therefore, a confidence interval $q_\alpha$ for $\bar P_1$
\be\label{eq:nwalpha}
nw_\alpha = F(q_\alpha)
\ee
is a one-sided confidence interval on $\bar N_W $, that is, with probability $1-\alpha$, the average Wigner-negativity as quantified by $\bar N_W$ is lower bounded by $nw_\alpha=F(q_\alpha)$. This can be used both in the calibration-independent $\hat{nw}_\alpha = F(\hat q_\alpha)$ and calibration-dependent $\hat{nw}^*_\alpha = F(\hat q^*_\alpha)$ settings.

\subsection{The crucial role of the single-mode hypothesis}
 \label{sec:multimode verify}

It is important to emphasize that the single-mode hypothesis is crucial in order to relate $P_1$ (or its estimated value  $\hat P_1(\bm p), \hat{P}_1^*(\bm p)$) to Wigner-negativity. In particular, the multi-mode single-photon state $\varrho$ of Eq.~\eqref{eq: single photon multimode} becomes Wigner-positive for $N>2$. In section \ref{sec:multimode generalzation}
we have discussed how the single-mode character of the emitted radiation can be verified in practice. Here we merely recall that a bound of the form $P_1^{[1]}\geq P_1-\varepsilon$, where $P^{[1]}_1$ is the single photon probability for a given mode, can be readily used to verify Wigner-negativity. One simply has 
\be\begin{split}
nw^{[1]}_\alpha &\geq F(q_\alpha - \varepsilon) \\
\text{p-value} &\leq \exp\left(-\frac{2 n \left(\min\{\bar X_T, \bar X_R\}-\varepsilon-\frac{1}{2}\right)^2}{16}\right),
\end{split}\ee
if $\min\{\bar X_T, \bar X_R\} > \frac{1}{2}+\varepsilon$, for Wigner-negativity of the said mode.

It is worth mentioning that a possibility to ensure that the detected radiation is single-mode is to perform homodyne measurements. In such a measurement the incoming beam is mixed with a strong local oscillator on a beam splitter, the intensity of the output beams are then measured with linear detectors and subtracted. Under the assumption that the local oscillator is single-mode\footnote{Note that the situation where $n$ modes $a_1,\dots,a_n$  are prepared in coherent states $\ket{\alpha_1} \ket{\alpha_2}\dots \ket{\alpha_n}$ with fixed phase relations, can be viewed as a coherent state of the mode $\bar a = \frac{\sum_k a_k \alpha_k}{\sqrt{\sum_k |\alpha_k|^2}}$, plus $n-1$ modes in the vacuum states.} the obtained signal is only sensitive to the single input mode identified by the local oscillator. Photon number statistics, and $P_1$ in particular, can be reconstructed from the statistics of a phase-averaged quadrature measurement~\cite{Munroe1995}, i.e. a homodyne measurement with a phase-randomized local oscillator.
This offers the possibility to use our bound on the Wigner-negativity $N_W(\rho)$ in Eq.~\eqref{eq: wigner measure} with measurements that are guaranteed to pick up a single mode.

\section{Experiment }
\label{sec: experiment}

To demonstrate the feasibility of our tools, we experimentally benchmark, witness and quantify the non-classical nature of a heralded single-photon source~\cite{Bruno2014} that is optimized for high efficiency of the heralded photon~\cite{Guerreiro2013}.
A periodically poled potassium titanyl phosphate (PPKTP) crystal is pumped by a Ti:Sapphire laser at $\lambda_p=\SI{771.8}{nm}$ in the picosecond pulsed regime with a repetition rate of \SI{76}{MHz} to create nondegenerate photon pairs at $\lambda_s=\SI{1541.3}{nm}$ (signal) and $\lambda_i=\SI{1546.1}{nm}$ (idler) via type-II spontaneous parametric down-conversion (SPDC).
The pair creation probability per pump pulse is set to $P_\mathrm{pair} \approx \num{1.0E-3}$ and high-purity heralded signal photons are ensured by spectrally filtering the heralding idler photons using a dense wavelength division multiplexer at ITU channel 39. From a joint spectral intensity measurement~\cite{Zielnicki2018}, we estimate the spectral purity of the heralded photon to be $98.59\% \pm 0.04\%$.
In this way, we herald signal photons at a rate of \SI{19.1}{kcps}.

For the heralded auto-correlation measurement, the signal photon is sent to a 50/50 fiber coupler (AFW FOBC). All photons are detected by MoSi superconducting nanowire single-photon detectors~\cite{Caloz2018} and time-correlated single-photon counting in a programmable time-to-digital converter (ID Quantique ID900) is used to register the detection events. Data are acquired for \SI{200}{s} in order to evaluate $\bm p = (p_{ \circ\circ},p_{\bullet \circ},p_{\circ \bullet},p_{\bullet\bullet})$ for the signal photons after the 50/50 beam splitter.

The dark count probabilities of the detectors we used for the auto-correlation measurement are fairly small, $p_\t{dc}\leq 4\times 10^{-7}$, and can be completely neglected. That is, the dark count corrections ($\approx 4\times 10^{-6}$) to the estimated values of the benchmark in Eqs.~(\ref{eq: corr dark counts},\ref{eq: dark count param}) are more than two orders of magnitude lower than the statistical noise, see Tables~\ref{tab:results} and \ref{tab:results_dependent}.

\subsection{Calibration-independent benchmark}

\begin{table}[b]
\capstart
\centering
\setlength\tabcolsep{10pt}
\begin{tabular}{c c c c} 
\toprule
$\eta_\text{s,tot}$ & $\hat{P}_1^T$ & $\hat{P}_1^R$ & $\hat{q}_{\alpha=10^{-10}}$ \\[0.5ex]
\midrule
62\,\% & \num{0.561(1)} & \num{0.678(1)} & 0.554 \\
52\,\% & \num{0.460(1)} & \num{0.573(1)} & 0.453 \\
42\,\% & \num{0.376(1)} & \num{0.465(1)} & 0.369 \\ 
\bottomrule
\end{tabular}
\caption{\label{tab:results} Results of the measurement for the three highest transmission efficiencies $\eta_\text{s,tot}$ of the heralded single-photon state. The values for $\hat{P}_1^T$ and $\hat{P}_1^R$ are calculated according to Eq.~\eqref{eq: general witness}. For the finite statistics analysis we calculate the confidence interval $\hat{q}_{\alpha}\leq \bar P_1$ from Eq.~(\ref{eq:q_alpha}) for the confidence level $1-\alpha= 1-\num{E-10}$.}
\end{table}

In a first step we apply our benchmark to the experimental results without taking the splitting ratio of the beam splitter and the detector efficiencies into account. The overall efficiency is 25\,\% for the heralding idler photons and 62\,\% for the heralded signal photons. In order to simulate a less efficient single-photon source, we introduce loss by inserting a fiber coupled variable attenuator (JDS Uniphase MV47W) into the heralded photon path before the 50/50 beam splitter and repeat the auto-correlation measurement for eight different transmission efficiencies $\eta_\text{att}$. Each transmission efficiency leads to a value for the pair $(\min(p_{\circ\_}, p_{\_\circ}),p_{\circ \circ})$ that is represented by a black cross in Fig.~\ref{fig:measurement}. In the same figure, we represent the polytope $\bm Q_{P}$ (defined after Eq.~\eqref{eq: p_o hat}) containing all possible values $(\min(p_{\circ\_}, p_{\_\circ}),p_{\circ \circ})$ associated to states  $\varrho$ with $\bra{1}\varrho\ket{1}\leq P$. Four polytopes are represented corresponding to the values $P\in\{0.25,0.5,0.75, 1\}$. A measurement result associated to a black cross lying outside a polytope $\bm Q_{P}$ is guaranteed to come from a state with a single-photon component satisfying $P_1 > P$.

For the measurements with the three highest transmission efficiencies, we give the results of our benchmark in Tab.~\ref{tab:results}. We conclude for the highest transmission for example, that the measured states have on average a single-photon component with a weight $\bar P_1 \geq 0.554$ with a confidence level of $1-10^{-10}$. 

\subsection{Calibration-dependent benchmark}

\begin{table}[t]
\capstart
\centering
\setlength\tabcolsep{5pt}
\begin{tabular}{l c c c l c c} 
\toprule
Mode 	& $\eta_\text{tot}$ & $\eta_c$ & $\eta_{f}$ & & $\eta_{t}$ & $\eta_{d}$ \\[0.5ex]
\midrule
Idler 			& 25\,\% & 80\,\% & 50\,\% & & 83\,\% & 75\,\% \\
\cmidrule{1-7}
\multirow{2}{*}{Signal} & \multirow{2}{*}{62\,\%} & \multirow{2}{*}{80\,\%} & \multirow{2}{*}{-} & R & 43\,\% & 92\,\% \\
\cmidrule{5-7}
& &  & & T & 44\,\% & 85\,\% \\
\bottomrule
\end{tabular}
\caption{\label{tab:efficiencies} Characterization of the loss for idler (heralding) and signal (heralded) modes. $\eta_\text{tot}$, total efficiency; $\eta_c$, fiber coupling efficiency; $\eta_{f}$, spectral filter transmission; $\eta_{t}$, fiber transmission including the insertion loss of the 50/50 fiber coupler, connectors and telecom fiber isolators for further pump rejection; $\eta_{d}$, detector efficiency.}
\end{table}

To compute the value of the calibration-dependent benchmark one needs to estimate the detector efficiencies and the reflection/transmission coefficient of the beam splitter. In order to characterize the detectors, we use the standard method, see e.g.~\cite{Caloz2018} for a detailed description. For our setup we find that the beam splitter coefficients are bounded by $\hat{r}\in[0.49,0.50]$ and the detector efficiencies are upper bounded by $(\hat{\eta}_R,\hat{\eta}_T)=(0.95,0.88)$.  The upper bounds for the detector efficiencies are obtained from results of the measured detection efficiencies given in Tab.~\ref{tab:efficiencies} by adding three times the measurement uncertainty of around 0.01, see Supplementary Material of~\cite{Caloz2018}.

Under the assumptions that the $(\hat r, \hat \eta_R, \hat \eta_T)$ belong to these intervals,  the values of $\hat{P}_1^{T\ast}(\bm p)$ and $\hat{P}_1^{R\ast}(\bm p)$ as measured in our experiment are given in Tab.~\ref{tab:results_dependent} for the three highest transmission efficiencies. The confidence interval $\hat{q}_{\alpha}^{\ast} \leq \bar P_1$  is also reported for a confidence level of $1-\alpha=1-$ \num{E-10}.

\begin{table}[ht]
\capstart
\centering
\setlength\tabcolsep{10pt}
\begin{tabular}{c c c c} 
\toprule
$\eta_\text{s,tot}$ & $\hat{P}_1^{T\ast}$ & $\hat{P}_1^{R\ast}$ & $\hat{q}_{\alpha=10^{-10}}^{\ast}$\\[0.5ex]
\midrule
62\,\% & \num{0.658(1)} & \num{0.683(1)} & 0.677 \\
52\,\% & \num{0.544(1)} & \num{0.573(1)} & 0.566 \\
42\,\% & \num{0.444(1)} & \num{0.466(1)} & 0.459 \\
\bottomrule
\end{tabular}
\caption{\label{tab:results_dependent} Results of the measurement including the imperfect beam splitter ratio with $(1-\hat t,\hat r)=(0.49,0.50)$ and the non-unit detection efficiencies by using the upper bounds $(\hat{\eta}_R,\hat{\eta}_T)=(0.95,0.88)$. The values for $\hat{P}_1^{T\ast}$ and $\hat{P}_1^{R\ast}$ are calculated with Eq.~\eqref{parambound}. The confidence intervals $\hat{q}_{\alpha}^{\ast}\leq \bar P_1$ in the finite statistics analysis are calculated for a confidence level of $1-\alpha=1-\num{E-10}$.}
\end{table}

\subsection{Non-classicality of the source}

\label{sec: exp wigner}

As already mentioned, there is no guarantee that the single-mode assumption is exactly satisfied in our experiment, hence we have to estimate the mode purity of the source. We assume that the spatial mode purity is guaranteed by coupling the photons into a single-mode fiber. The polarization-purity is ensured by the fact that the signal and idler photons are separated with a polarizing beam splitter. To estimate the spectral purity, we apply the standard approach that relies on a physical model of the source, which we believe to properly describe the experiment. Precisely, we assume that the SPDC process responsible for the generation of the photon pairs is of the form 
\be
H_\t{SPDC} \propto \int \dd w_s \dd w_i f(w_s,w_i) a^\dag(w_s) b^\dag(w_i), +\text{h.c.}, 
\ee
where $a(w_s)$ and $b(w_i)$ are the frequency field-modes of the signal/idler photons with $[a(w),a^\dag(w')] = [b(w),b^\dag(w')] =\delta(w-w')$. At low pumping power, we reconstruct $f(w_s,w_i)$ with a signal-idler joint spectral intensity measurement. Via a 2D-Gaussian fit we perform a singular value decomposition of $f(w_s,w_i)$ to rewrite the interaction in the form
\be
H_\t{SPDC} \propto  \sum_k \sqrt{\lambda_k} a_k^\dag b_k^\dag +\text{h.c.},
\ee
where $[a_k,a_\ell^\dag]=[b_k,b_\ell^\dag]=\delta_{k\ell}$ now describe discrete spectral modes. With this procedure the largest Schmidt coefficient $\lambda_1$ is computed to be $\lambda_1=\num{0.99292(18)}$, where the standard deviation  $\sigma_{\lambda_1}=1.8\times10^{-4}$ is obtained from a Monte Carlo method assuming Poissonian count statistics in the joint spectral intensity measurement. 
Therefore, assuming that the efficiency of the trigger detector is the same for all idler modes, we obtain the leading order estimate  
\be
1-\varepsilon \approx \lambda_1 - 3\,  \sigma_{\lambda_1} = 99.24\%,
\ee
of the probability that the modes $a_{k\geq2}$ are empty conditional to the detection of an idler photon. This corresponds to the mode-purity of $\sum_k \lambda_k^2 \approx 98.59\%$. 

With the help of Eq.~\eqref{eq: multi-mode P} we can take this into account for the quantification of Wigner-negativity, resulting in $\hat{nw}_{\alpha=10^{-10}}=0.0046$ for the case of no added loss on the heralded single-photon state and no assumptions on the calibration of the measurement apparatus. The corresponding p-value and the results for the measurement-apparatus-dependent case are given in Tab.~\ref{tab:results_wigner}.

\begin{table}[ht]
\capstart
\centering
\setlength\tabcolsep{4pt}
\begin{tabular}{c c c c c} 
\toprule
$\eta_\text{s,tot}$ & $\hat{nw}_{\alpha=10^{-10}}$ & p-value & $\hat{nw}_{\alpha=10^{-10}}^{\ast}$ & p-value$^\ast$ \\[0.5ex]
\midrule
62\,\% & 0.0046 & \num{E-603} & 0.053 & \num{E-6420} \\
52\,\% & 0 & $\cross$ & 0.0072 & \num{E-894} \\
42\,\% & 0 & $\cross$ & 0 & $\cross$ \\
\bottomrule
\end{tabular}
\caption{\label{tab:results_wigner} Wigner-negativity in our experiment for the three highest transmission efficiencies $\eta_\text{s,tot}$ of the heralded single-photon state. The confidence interval on the measure of Wigner-negativity $\hat{nw}_{\alpha}\leq \bar N_W$ is obtained from Eq.~(\ref{eq:nwalpha}) for the confidence level $\alpha= \num{E-10}$, assuming the reduced single-mode $P_1^{[1]}$ as given in Eq.~(\ref{eq: multi-mode P}). Further, we give the p-value according to Eq.~(\ref{pvalue}) associated with the hypothesis that the measured states are on average Wigner-positive. The quantities with a $^*$ are taking the detector efficiencies into account and are obtained accordingly from Eqs.~(\ref{parambound}) and (\ref{eq: multi-mode P}).}
\end{table}

\section{Conclusion}
Auto-correlation measurements are commonly used to assess the quality and the quantum nature of single-photon sources, that is, they are used to check that a given source does not emit more than one photon and its emission is non-classical in the sense that its P-distribution is non-positive or that its state is non-Gaussian. We have shown that the statistics obtained from these measurements is actually richer. They can be used to lower bound the probability that a given source actually produces a single photon. We argued that this probability is a good benchmark for single-photon sources as it captures both its quality and its efficiency. Moreover, we showed that if the mode purity of the source can be assessed the lower bound on the single-photon emission probability can be used to witness and quantify the negativity of the Wigner function, a stronger form of non-classicality than the negativity of the P-distribution and the non-Gaussianity. We have proposed practical tools to benchmark single-photon sources and characterize its Wigner-negativity this way. With this material in hand, we hope that the community which is developing single-photon sources could exploit the statistics of their auto-correlation measurements in a more enlightening way. 

\begin{acknowledgments}
We thank R.J. Warburton for fruitful discussions at an early stage of the project. This work was supported by the Swiss National Science Foundation (SNSF) under Grant No. 200020-182664. E.O
acknowledges support from the Government of Spain
(FIS2020-TRANQI and Severo Ochoa CEX2019-000910-
S), Fundació Cellex, Fundació Mir-Puig, Generalitat de
Catalunya (CERCA, AGAUR SGR 1381) and from the
ERC AdGCERQUT. 
\end{acknowledgments}

\section*{Data availability}
The data supporting the experimental results within this paper are available on the Zenodo data repository~\cite{ZenodoRepository}.

\appendix

\section{Auto-correlation function and auto-correlation measurement}
\label{app: autocorrelation}

Here we briefly recall the definitions of the auto-correlation function $g^{(2)}$. We will use the notation introduced in Section~\ref{sec: apparatus}, \ref{sec: benchmark} and \ref{sec: benchmark ideal} of the main text, and assume that the beam splitter is balanced and the two detectors have equal efficiency, i.e. $t=r=\frac{1}{2}$ in Eq.~\eqref{eq: POVM elements}.

Historically~\cite{scully1999quantum}, the auto-correlation function was defined as the ratio 
\be
g^{(2)}= \frac{\mean{a^{\dag2}a^2}}{\mean{a^\dag a}^2}=\frac{\mean{(a^\dag a)^2}-\mean{a^\dag a}}{\mean{a^\dag a}^2}
\ee
and the efficiency of the source can be characterized by the average number of photons it emits $I=\mean{a^\dag a}$. Then it is not difficult to see that the bound 
\be
P_1 \geq 2 \mean{a^\dag a}- \mean{(a^\dag a)^2} = I  - I^2 g^{(2)} 
\ee
is a tight benchmark. To see this note that in the $\big(\mean{a^\dag a},\mean{(a^\dag a)^2}\big)$ plane the quantity $w= 2 \mean{a^\dag a}- \mean{(a^\dag a)^2}$ measures the distance from the line connecting the points $(0,0)$ and $(2,4)$ corresponding to Fock states $\ket{0}$ and $\ket{2}$. With $w=1$ for the point $(1,1)$ corresponding to the single-photon state $\ket{1}$.
In practice, one can not directly measure $\mean{(a^\dag a)^2}$ and $\mean{a^\dag a}$, but can approximate these values by increasing the loss artificially, as $E_{\bullet\_} = \frac{a^\dag a}{2}   \, \eta + O(\eta^2)$  and $E_{\bullet\bullet} = \frac{a^{\dag 2}a^2}{4} \eta^2 +O(\eta^3)$ for $\frac{1}{\eta}\gg a^\dag a$ (recall that $E_\bullet\_= E_{\bullet\bullet}+E_{\bullet\circ}$). Such an approach thus requires a precise control of the efficiency $\eta$ and is statistically inefficient, since additional losses are introduced. 

Alternatively~\cite{sekatski2012detector}, the auto-correlation function can by directly defined as 
\be
\tilde{g}^{(2)} =\frac{\mean{E_{\bullet \bullet}}}{\mean{E_{\bullet \_}}\mean{E_{\_ \bullet}}}.
\ee
The efficiency can also be characterized by probability that a source produces a click $\tilde{I}=\mean{E_{\bullet\bullet}+E_{\bullet\circ}+E_{\circ\bullet}}$.  Given the two values $\tilde{g}^{(2)}$ and $\tilde{I}$ it is then possible to reconstruct the full probability distribution $\bm p=(\mean{E_{\circ\circ}},\mean{E_{\bullet\circ}},\mean{E_{\circ\bullet}},\mean{E_{\bullet\bullet}})$, since we assumed $\mean{E_{\bullet\circ}}= \mean{E_{\circ\bullet}}$ so that $\bf p$ is defined by two parameters. As argued above the two functions coincide $g^{(2)}=\tilde{g}^{(2)}$ in the limit $\eta\to 0$.

In both cases the auto-correlation measurements relies on the setup of the Fig.~\ref{fig:schemeg2}. The measurement data can thus be readily used to estimate $\bm p$ and compute the benchmarks proposed in this paper.

To finish the discussion of the auto-correlation functions we recall that both $g^{(2)}$ and $\tilde g^{(2)}$ are witnesses of the non-classicality of the state $\rho$~\cite{sekatski2012detector}, i.e. $g^{(2)}, \tilde g^{(2)}<1$ is only possible for states whose P-function admits negative values. In fact, this is true in a more general context, as given by the following observation.\\

\noindent \textbf{Observation.} \textit{For any two binary POVMs $\{M_\bullet,M_\circ\}$ and $\{M'_\bullet, M'_\circ\}$ measured at the two outputs of a beam splitter (Fig.~\ref{fig:schemeg2}), the inequality}
\be
G^{(2)} = \frac{\mean{M_\bullet\otimes M'_\bullet}}{\mean{M_\bullet\otimes \id} \mean{\id \otimes M'_\bullet}}<1
\ee
\textit{is a witness of non-classicality (P-function taking negative values), as long as the POVM element of individual detectors are only functions of the number of photons $\bra{n} M_\bullet \ket{m} =  \delta_{n,m} p_\bullet(\ket{n})$ and the click probabilities are increasing functions of the photon number $p_\bullet(\ket{n}) \geq p_\bullet(\ket{m})$ for $n\geq m$ (and the same for $M_\bullet'$). }\\

For the sake of completeness we prove the above statement here. A coherent state splits into two coherent states on a beam splitter $\ket{\alpha} \mapsto_{BS} \ket{\sqrt{r}\alpha}\ket{\sqrt{t}\alpha}$. Hence for a coherent state 
\be\begin{split}
\mean{M_\bullet\otimes M'_\bullet} &= \bra{\sqrt{r}\alpha}M_\bullet \ket{\sqrt{r}\alpha} \bra{\sqrt{t}\alpha}M'_\bullet \ket{\sqrt{t}\alpha}\\
&=p_\bullet(\sqrt{r |\alpha|^2})\,  p_\bullet'(\sqrt{t|\alpha|^2})  \\
&= \mean{M_\bullet\otimes \id} \mean{\id \otimes M'_\bullet}    
\end{split}
\ee
and $G^{(2)}=1$. Then for any mixture of coherent states $\rho_{cl}=\int d^2 \alpha \text{P}(\alpha) \ketbra{\alpha}$ one finds
\be
    \mean{M_\bullet\otimes M'_\bullet} = \int d^2 \alpha \text{P}(\alpha)  p_\bullet(\sqrt{r |\alpha|^2}) p_\bullet'(\sqrt{t|\alpha|^2}).
\ee
To shorten the equations let us denote $z=|\alpha|$, $f(z)=p_\bullet(\sqrt{r |\alpha|^2})$,  $g(z)=p'_\bullet(\sqrt{t |\alpha|^2})$, $\mu(z) = 2z \int \dd \varphi \text{P}\left(z e^{\ii \varphi}\right)$ and $\dd \mu(z) = \dd z \mu(z)$ with $\int \dd \mu(z)=1$, such that

\be\begin{split}
 \mean{M_\bullet\otimes M'_\bullet} &= \int \dd\mu(z)f(z) g(z) \\
    & \!\!\! \!\!\!= \int \dd \mu(z) \dd \mu(z') \frac{1}{2} (f(z)g(z)+f(z')g(z')),\\
   \mean{M_\bullet\otimes \id} &\mean{\id \otimes M'_\bullet} = \int \dd \mu(z)f(z)\int \dd \mu(z')g(z') \\
    & \!\!\! \!\!\!=  \int \dd \mu(z) \dd \mu(z')\,  \frac{1}{2} (f(z) g(z')+f(z') g(z)).
\end{split} \ee

Without loss of generality consider $z'\geq z$, by $p_\bullet(\ket{n}) \geq p_\bullet(\ket{m})$ for $n\geq m$ it follows that $f(z')\geq f(z)$ for $z'\geq z$ (using the fact that a coherent state has a Poissonian photon number distribution). Hence, one can write
\be\begin{split}
f(z') = f(z) +\Delta_f \qquad \Delta_f\geq 0\\
g(z') = g(z) +\Delta_g \qquad \Delta_g\geq 0
\end{split}
\ee
and 
\be
\begin{split}
    &(f(z) g(z)+f(z') g(z')) - (f(z)g(z')+f(z')g(z)) \\
    &= \Delta_f \Delta_g 
     \geq 0
\end{split}
\ee
for each $z$ and $z'$. Therefore the inequality is also true for the integrals
\be
\mean{M_\bullet\otimes M'_\bullet} \geq \mean{M_\bullet\otimes \id} \mean{\id \otimes M'_\bullet}.
\ee
Showing that $G^{(2)}\geq 1 $ for any state with a positive P-function.
Notably, the same proof works if both $\bra{n}M_\bullet\ket{n}$ and $\bra{n}M'_\bullet\ket{n}$ are
decreasing functions of $n$, e.g. if they are replaced with $M_\circ$ and $M_\circ'$.  The above observation has been used to propose an experiment where non-classicality of light would be 
demonstrated by directly using human eyes as detectors~\cite{Dodel2017proposalwitnessing}.

\section{POVM corresponding to photon detection preceded by a beam splitter}

\label{sec:povm}
Since two single-photon detectors after a beam splitter cannot detect any coherence between Fock states of an incoming state, we can  consider a state $\rho=\sum_n \frac{c_n}{n!} (a^{\dag})^n\ket{0}\bra{0}a^n$ without loss of generality. The state $\rho$ arrives at a beam splitter with transmittance t, the resulting state is (ignoring again the coherence between the Fock states)
\be\begin{split}
\rho_r&=\sum_{n,k}c_n\binom{n}{k}(t^k)(1-t)^{n-k}\ket{k,n-k}\bra{k,n-k}.\end{split}\ee
One can then compute the probability of the different events for the state  $\rho_r$, for example 
\be\begin{split} 
P_{\circ \bullet} &=\text{Tr}(\text{E}_{\circ}\otimes\text{E}_{\bullet}\rho_r)=\text{Tr}(\text{E}_{\circ}\rho_r)-\text{Tr}(\text{E}_{\circ}\otimes\text{E}_{\circ}\rho_r) 
\end{split}\ee
where
\be\begin{split} 
\nonumber
\text{Tr}(\text{E}_{\circ}\otimes\text{E}_{\circ}\rho_r)&=\sum_{n}c_n^2(1-\eta)^n\sum_{k=0}^n \binom{n}{k}t^k(1-t)^{n-k}\\
&=\sum_{n}c_n^2(1-\eta)^n\\
&=\text{Tr}((1-\eta)^{a^{\dag}a} \rho)
\end{split}\ee
and
\be\begin{split} 
\nonumber
\text{Tr}(\text{E}_{\circ}\rho_r)&=\sum_{n}c_n^2\sum_{k=0}^n \binom{n}{k}t^k(1-t)^{n-k}(1-\eta)^k\\
&=\sum_{n}c_n^2(1-t\eta)^n\\
&=\text{Tr}((1-t\eta)^{a^{\dag}a} \rho).
\end{split}\ee

The resulting POVM is thus $\text{E}_{\circ\bullet}=(1-t\eta)^{a^{\dag}a}-(1-\eta)^{a^{\dag}a}.$ 

Let us consider a two-mode state (generalization to $n$ modes is straightforward) $\rho=\sum_{n_1,n_2} \frac{c_{n_1,n_2}}{n_1!n_2!} (a_1^{\dag})^{n_1} (a_2^{\dag})^{n_2}\ket{0}\bra{0}a_1^{n_1} a_2^{n_2}.$ The POVM element corresponding to no-clicks after the beam splitter on each mode now reads $\text{E}^{(2)}_{\circ}=\text{E}_{\circ}\otimes\text{E}_{\circ}.$ The state after the beam splitter is now
\be\begin{split} 
&\rho_r=\sum_{n_1,k_1,n_2,k_2}C_{n_1,k_1,n_2,k_2}\\
&\ket{k_1,n_1-k_1,k_2,n_2-k_2}\bra{k_1,n_1-k_1,k_2,n_2-k_2}
\end{split}\ee
where $C_{n_1,k_1,n_2,k_2}=\binom{n_1}{k_1}\binom{n_2}{k_2}t^{k_1+k_2}(1-t)^{n_1+n_2-k_1-k_2}.$ We focus on the event click on the first detector and no click on the second
\be\begin{split} 
\nonumber
&\text{Tr}(\text{E}^{2}_{\circ}\otimes\text{E}_{\circ}^{2}\rho_r)\\
&=\sum_{n_1,n_2}c_{n_1,n_2}^2(1-\eta)^{n_1+n_2}\\
&\sum_{k_1,k_2=0}^{n_1,n_2} \binom{n_2}{k_2}\binom{n_1}{k_1}t^{k_1+k_2}(1-t)^{n_1+n_2-k_1-k_2}\\
&=\sum_{n}c_n^2(1-\eta)^{n_1+n_2}\\
&=\text{Tr}((1-\eta)^{\hat{n}} \rho)
\end{split}\ee
similarly
\be
\text{Tr}(\text{E}^{2}_{\circ}\rho_r)=\text{Tr}((1-t\eta)^{\hat{n}} \rho).
\ee
We retrieve that the POVM  elements in the multi-mode case are given by 
\be\begin{split}
E_{\circ \circ} &= (1-\eta)^{\hat{n}} \\
E_{\bullet \circ} &= (1-\eta\, t)^{\hat{n}}- (1-\eta)^{\hat{n}}\\
E_{\circ \bullet} &= (1-\eta\, r)^{\hat{n}}- (1-\eta)^{\hat{n}}\\
E_{\bullet \bullet} & = \id - E_{\circ \circ} -E_{\bullet \circ} - E_{\circ \bullet}.
\end{split}
\ee

\section{The effect of loss on \texorpdfstring{$P_1$}{P1}}
\label{sec:loss}

We show here that the set of states $\rho$ with $P_1\geq \frac{2}{3}$ is closed under losses. 
Consider a state $\rho$ associated with a single-photon component $P_1=P$. Let us apply infinitesimal transmission losses $\eta=1-\dd \epsilon$. After the loss, the photon number distribution $P_n=\langle n|\rho|n\rangle$ reads
\be
P_n(\epsilon) = (1- n\dd\epsilon)P_n + \dd \epsilon (n+1)P_{n+1}.
\ee
In particular, 
\be\begin{split}
\frac{\dd}{\dd \epsilon} P_1 &= -P_1 +2 P_2\\
&\leq -P + 2(1-P)\\
 & =  2 - 3 P
\end{split}
\ee
which is negative for $P\geq 2/3.$ 

\begin{figure}
    \centering
    \includegraphics[width=\columnwidth]{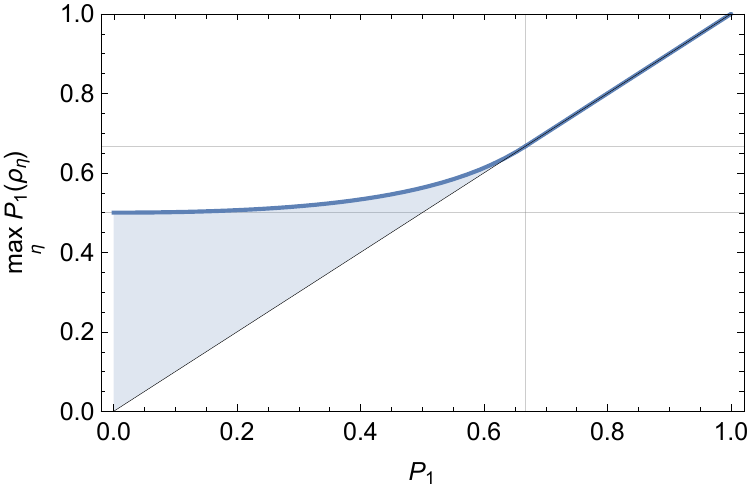}
    \caption{Consider the sate $\rho = P_1\ketbra{1}{1} + (1-P_1)\ketbra{2}$, which is transformed intro $\rho_\eta$ by a loss channel of transmission $\eta$. Fro any $P_1$ there is a value of $\eta$ which maximize the single-photon probability $P_1(\rho_\eta)$ after losses, see Eq.~\eqref{eq: max P1 loss}. The blue line depicts $\max_\eta P_1(\rho_\eta)$ as the function $P_1$, and the black line is simply $P_1$. One sees that for this example the single-photon component can be increased by losses if $P_1<2/3$, as illustrated by the shaded area. Conversely, we show that for any single-mode state $\rho'$ with $P_1\geq 2/3,$ the single-photon component can not be increased by losses.}
    \label{fig:P1}
\end{figure}

On the other hand, there are states with $P_1<2/3$ for which the single-photon probability can be increased substantially by losses. Consider a channel with transmission efficiency $\eta$ and apply it to the state
$\rho = P_1 \prjct{1}+(1-P_1)\prjct{2}$. 
This leads to a state $\rho_\eta$ having a single-photon component
\be
P_1(\rho_\eta) = \eta P_1 + 2\eta(1-\eta)(1-P_1).
\ee
This quantity is maximized at
\be\label{eq: max P1 loss}
\max_\eta P_1(\rho_\eta) = \begin{cases}
\frac{(2-P_1)^2}{8 (1-P_1)} & P_1 \leq \frac{2}{3}\\
P_1 & P_1>\frac{2}{3},
\end{cases}
\ee
the maximum being depicted in Fig.~\ref{fig:P1} as a function of $P_1$. To give a concrete example, for the initial $\rho =\frac{1}{2}(\prjct{1}+\prjct{2})$, the weight of the single-photon component can be increased to $P_1(\rho_{\eta=3/4}) = 1/2 + 1/16 = 0.562$ while for the Fock state $\rho =\prjct{2}$, it is possible to reach $P_1(\rho_{\eta=1/2}) = 1/2$.

\section{Parameter dependent witness} 
\label{sec:param}

We consider the case where the two detectors have efficiencies $\eta_T$ and $\eta_R$ and the beam splitter has reflectance $r$ and transmittance $t$ (with $t+r=1$). For an incoming Fock state $\ket{n}$ the probabilities of clicks are given by

\be\begin{split}
   f_n = p_{\circ\bullet }^{\ket{n}} &= (1-\eta_R r)^n-(1- \eta_T t -  \eta_R r)^n\\
    h_n =p_{\bullet\circ }^{\ket{n}} &= (1-\eta_T t)^n-(1-\eta_R r-\eta_T t)^n\\
    g_n= p_{\bullet\bullet }^{\ket{n}} &=1+(1-\eta_R r - \eta_T t)^n+\\
    &-(1-\eta_R r)^n-(1-\eta_T t)^n.
   \end{split}
\ee

For a mixture of Fock states $\rho = \sum_n P_n \prjct{n}$ one has
\be
p_{\circ\bullet }=\sum_{n=1}^{\infty}P_n f_n,
\ee
from which we get
\be\label{eq: P1 bound D3}
    P_1 = \frac{1}{f_1}\left(p_{ \circ \bullet} - \sum_{n\geq 2} P_n f_n\right).
\ee    
In order to set a lower bound on $P_1$ we thus need to upper bound the term $ \sum_{n\geq 2} P_n f_n$. We also have 
\be
    p_{\bullet\bullet } = \sum_{n\geq 2}P_n g_n.
\ee
Therefore, to derive a benchmark for $P_1$ we are looking for a function $f_*(p_{\bullet\bullet })$ such that
\be\label{eq: maximization f*}
\begin{split}
f_*(p_{\bullet\bullet }) &= \max_{\bf p>0} \sum_{n\geq 2} P_n f_n \\
        & \t{s.t.} \sum_{n\geq 2} P_n g_n = p_{\bullet\bullet } 
\end{split}
\ee
To find $f_*(p_{\bullet\bullet })$ we define $q_n =  P_n g_n$, so that the maximization can be rewritten as
\be\begin{split}
    f_*(p_{\bullet\bullet }) =&\max  \sum_{n\geq 2} q_n \frac{f_n}{g_n} \\
    &\t{s.t.}\, \sum_{n\geq 2} q_n =  p_{\bullet\bullet }.
\end{split}
\ee
Using $\sum_n q_n \frac{f_n}{g_n} \leq (\sum_n q_n) \left(\max_{n}\frac{f_n}{g_n}\right)$ we see that the solution of Eq.~\eqref{eq: maximization f*} satisfies
\be\label{eq: fn/gn}
f_*(p_{\bullet\bullet }) \leq  p_{\bullet\bullet } \left(\max_{n\geq 2} \frac{f_n}{g_n}\right).
\ee

We prove right after that the maximum is achieved for $n=2$. By plugging $\sum_{n\geq 2} P_n f_n \leq f_*(p_{\bullet\bullet }) \leq p_{\bullet\bullet } \frac{f_2}{g_2}$ in Eq.~\eqref{eq: P1 bound D3}, we get the desired inequality
\be
 P_1 \geq \frac{1}{f_1}\left(p_{\circ\bullet } - p_{\bullet\bullet } \frac{f_2}{g_2}\right).
\ee
The same bound holds with $p_{\bullet \circ}$ instead of $p_{\circ \bullet}$  and $h_n$ instead of $f_n$. The right-hand side of these two inequalities are a function of the observed probabilities $\bm p$ and a lower bound on $P_1$, hence defining two benchmarks
\be
\begin{split}
    \hat{P}_1^{T*}(\bm p) &= \frac{1}{f_1} p_{\circ\bullet } -  \frac{f_2}{f_1 g_2}p_{\bullet\bullet },\\
    \hat{P}_1^{R*}(\bm p) &= \frac{1}{h_1} p_{\bullet \circ} -  \frac{h_2}{h_1 g_2}p_{\bullet\bullet }.
\end{split}
\ee
The best option is to consider the larger value of $p_{\bullet \circ}$ and $p_{\circ\bullet }$.

\smallskip

\textit{The proof of $\max_n \frac{f_n}{g_n} =\frac{f_2}{g_2}$.} To maximize the ratio $\frac{f_n}{g_n}$ express it as
\be\begin{split}
 \frac{f_n}{g_n}  &= \frac{(1-\eta_R r)^n-(1- \eta_T t -  \eta_R r)^n}{1-(1-\eta_T t)^n-(1-\eta_R r)^n + (1-\eta_R r - \eta_T t)^n
} 
 \\ &= \frac{1}{\frac{1-(1-\eta_T t)^n}{(1-\eta_R r)^n-(1-\eta_R r-\eta_T t)^n}-1}.
\end{split}
\ee
 Manifestly, maximizing  $\frac{f_n}{g_n}$  is equivalent to minimizing $\frac{1-(1-\eta_T t)^n}{(1-\eta_R r)^n-(1-\eta_R r-\eta_T t)^n}$. In other words we want to show that for $n\geq 2$ the fraction
 \be
 \frac{1^n-(1-x)^n}{y^n - (y-x)^n},
 \ee 
 with $x=\eta_T t$ and $y=1-\eta_R r$ satisfying $0<x<y<1$, is minimized at $n=2$. However, it is enough to show that the expression $ 
 \frac{1^n-(1-x)^n}{y^n - (y-x)^n}$ is increasing with $n$. To do so, let us derive this quantity with respect to $n$. We have
\be\begin{split}
&\frac{\dd }{\dd n} \frac{1-(1-x)^n}{y^n - (y-x)^n} = \frac{1}{(y^n -(x-y)^n)^2} \\
&\times\Big((1-x)^n-1) \left(y^n \log (y)-(y-x)^n \log (y-x)\right) \\
&-(1-x)^n \log (1-x) (y^n-(y-x)^n )\Big) .  
\end{split}
\ee
To show that it is positive we can omit the denominator $(y^n -(x-y)^n)^2$. 
Labeling $a=(1-x)^n, b = y^n, c= (y-x)^n$ and noting that $\log(x^\frac{1}{n})= \frac{1}{n}\log(x)$ we get
\be
\nonumber
\frac{\dd }{\dd n} \frac{1-(1-x)^n}{y^n - (y-x)^n} \geq 0 \Longleftrightarrow f(a,b,c) \geq 0,
\ee
with 
\be
f(a,b,c) = (a -1 ) (b \log(b) - c \log(c)) - a \log(a)(b-c). 
\ee
It remains to show that the function $f(a,b,c)$ is positive for $a,b > c$. Note that it is a decreasing function of $c$, as
\be\label{eq: temptemp}\begin{split}
\frac{\dd}{\dd c} f(a,b,c) &= (1-a) (\log (c)+1)+ a \log (a) \\
    &\leq(1-a) (\log (a)+1)+ a \log (a) \\
    & = 1 - a +\log(a) \\
    & \leq 0
\end{split}
\ee
using a standard inequality for the logarithm $\log(a)\leq 1-a$. We can thus only verify the positivity of the function for the maximal possible value of $c$. There are, however, two possibilities
$a\geq b$ and $b>a$. For $a \geq b$ we set $c=b$ and obtain
\be
f(a,b,c) \geq f(a,b,b) = 0. 
\ee
For $b>a$ we set $c=a$ and get
\be\begin{split}
f(a,b,c) &\geq f(a,b,a) \\
& = (1- b) a \log (a)- (1-a) b \log (b).
\end{split}
\ee
To show that the last expression is positive, we divide it by $(1-a)( 1-b)$ to get 
\be
 \frac{a}{1-a} \log (a) -\frac{b}{1-b} \log (b),
\ee
and note that the function $ \frac{x}{1-x} \log (x)$ is decreasing ($\frac{\dd}{\dd x} \frac{x}{1-x} \log (x) =\frac{1-x +\log(x)}{(1-x)^2}\leq 0$ by Eq.~\eqref{eq: temptemp}). Therefore, $b\geq a$ implies 
\be
\frac{a}{1-a} \log (a) -\frac{b}{1-b} \log (b) \geq 0 \implies f(a,b,c)\geq 0.
\ee
Hence, the fraction $ \frac{1-(1-x)^n}{y^n - (y-x)^n}$ is increasing with $n$ and attains its minimum at the boundary $n=2$ of the interval $[2,\infty)$. Therefore, $\frac{f_n}{g_n}$ is maximized at $n=2$, which concludes the proof.

\section{Wigner-negativity measure}
\label{sec:wigneg}

For a single-mode state $\rho$, the Wigner function $W_\rho(\beta)$ is a quasi-probability distribution satisfying $\int \dd \beta^2 \, W_\rho(\beta)=1$. The negativity of the Wigner function ($W(\beta)<0$ for some $\beta \in \mathds{C}$) is an important non-classical feature of the state, as argued in the main text. A natural way to quantify this negativity is to measure the total quasi-probability where the function $W_\rho$ takes negative values, that is to compute
\be
N_W(\rho)= \int \dd\beta^2\frac{|W_\rho(\beta)| -W_\rho(\beta)}{2}.
\ee
This intuitive quantity was introduced in \cite{kenfack2004negativity}.
We now show that $N_W(\rho)$ is a good "measure" of Wigner negativity in the sense that it can not be increased by Gaussian operations. 

Pure Gaussian operations are displacements $\t{D}_\gamma=e^{\gamma a^\dag - \gamma^* a}$, single-mode squeezing $\t{SMS}_g =e^{\frac{g}{2}(a^{\dag 2} - a^2)} $, phase rotations $e^{i \varphi a^\dag a}$, or combination thereof. Consider a single-mode state $\rho$ with its Wigner function $W_\rho(\beta)$ and its Wigner negativity measure $N_W(\rho)$.  The effect of a displacement $\varrho = \t{D}_\gamma \rho \t{D}_\gamma^\dag$ on the Wigner function is a mere translation in phase space $W_\varrho(\beta) = W_\rho (\beta-\gamma)$, which does not affect the Wigner negativity measure $N_W(\rho)=N_W(\varrho)$. The same goes for a phase rotation, which merely transform $W_\varrho(\beta)=W_\rho(\beta e^{\ii \varphi})$.
For a squeezing operation $\varrho = \t{SMS}_g \rho \t{SMS}_g^\dag$, the Wigner function is transformed as
\begin{equation}
    W_\varrho(\beta) = W_\rho(\tilde \beta )
\end{equation}
where $\beta = \beta'+\ii \beta''$ and $\tilde\beta = e^g \beta' + e^{-g}  \ii \beta''$. This implies for the Wigner negativity measure that
\begin{equation}\begin{split}
    N_W(\varrho) &= \frac{1}{2} \int \dd \beta^2 \left(|W_\varrho(\beta)| - W_\varrho(\beta)\right)\\
    &=\frac{1}{2} \int \dd \beta'\, \dd \beta'' \left(|W_\rho(\tilde \beta)| - W_\rho(\tilde \beta)\right) \\
    &=\frac{1}{2} \int e^{-g}  \dd \tilde \beta'\, e^g \dd \tilde \beta'' \left(|W_\rho(\tilde \beta)| - W_\rho(\tilde \beta)\right)\\
    &=\frac{1}{2} \int \dd \tilde \beta^2 \left(|W_\rho(\tilde \beta)| - W_\rho(\tilde \beta)\right) \\
    & = N_W(\rho).
\end{split}
\end{equation}
Hence, $N_W(\rho)$ is also unchanged by squeezing. Moreover, the quantity
\begin{equation}
    N_W(p_1\rho_1 + p_2\rho_2) \leq p_1 N_W(\rho_1)+ p_2 N_W(\rho_2)
\end{equation}
is manifestly convex as $|p_1 W_1(\beta) + p_2 W_2(\beta)|\leq p_1| W_1(\beta)| + p_2 |W_2(\beta)|$. Hence, $N_W(\rho)$ is non-increasing  under mixtures of pure Gaussian operations. We conclude that $N_W(\rho)$ is a reasonable measure of Wigner negativity.

\smallskip

Let us now show how the Wigner negativity measure of a given state can be related to the weight of its single-photon component. The Wigner function for an arbitrary Fock state $|n\rangle$ reads~\cite{vogel2006quantum}
\be
W_{n}(\beta)=\frac{2(-1)^n}{\pi} e^{-2|\beta|^2} L_n\left(4|\beta|^2)\right),
\ee
with $|W_{n}(\beta)|\leq \frac{2}{\pi}$ since $e^{-x/2}|L_n(x)|\leq 1$ ($L_n$ are Laguerre polynomials).
Hence, for any mixture of Fock states $\rho = \sum P_n \prjct{n}$, we have
\be
\begin{split}
W_\rho (\beta)&= P_1 W_{1}(\beta) + \sum_{n\neq 1} P_n W_{n}(\beta)\\
    &\leq  - \frac{2}{\pi}\left( P_1 (1- 4|\beta|^2) e^{-2|\beta|^2} - (1-P_1)\right).
\end{split}
\ee
\begin{figure}
    \centering
    \includegraphics[width=\columnwidth]{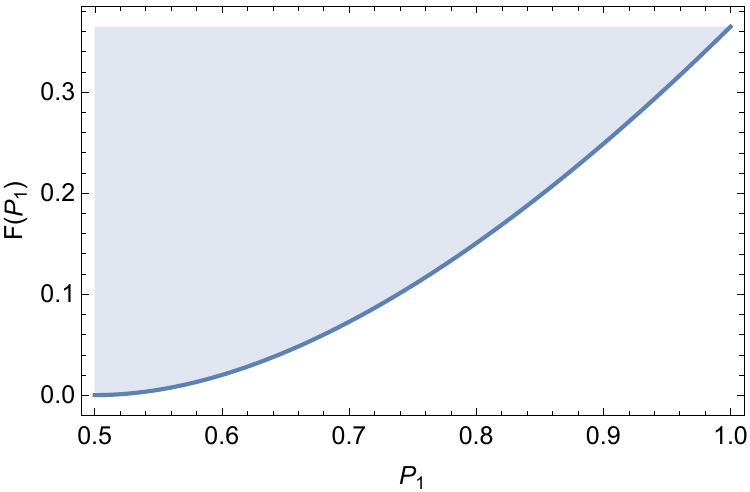}
    \caption{A sketch of the function $F(P_1)$ in Eq.~\eqref{eq: LB negativity} that lower bounds the Wigner-negativity of a state $N_W(\rho)$, for $P_1 =\bra{1}\rho\ket{1} \in[0.5,1]$. Any single-mode state $\rho$ with single-photon probability $P_1$ (or higher) has Wigner-negativity $N_W(\rho)$ above the blue line, as illustrated by the shaded area. }
    \label{fig: int neg}
\end{figure}

For $P_1\geq 1/2$, the Wigner function is negative in the phase-space region with
\be\label{eq: W negative}\begin{split}
    (1- 4|\beta|^2) e^{-2|\beta|^2} &>\frac{1-P_1}{P_1} \Longleftrightarrow\\
   |\beta|^2&< \underbrace{\frac{1}{4}\left(1 - 2 \, w_0\left( \frac{\sqrt{e}}{2}\frac{1-P_1}{P_1}\right)\right)}_{\equiv \ell(P_1)}, 
\end{split}
\ee
where $w_0(x)$ is the principal branch of the Lambert W function. Eq.~\eqref{eq: W negative} defines a disk 
\be
\t{disk}(P_1)=\Big \{ \beta\in \mathds{C}\Big|\, |\beta|^2 \leq \ell(P_1)\Big \}
\ee
in phase-space centered at the origin, where the Wigner function is negative. 
We can now  compute the integral over this region
\be\label{eq: LB negativity}\begin{split}
N_W(\rho) &\geq   \int_{\t{disk}(P_1)} \dd \beta^2 |W_\rho(\beta)| \\
& =  \int_0^{\sqrt{\ell(P_1)}} 2 \pi r \dd r \, |W_\rho(r)|\\ 
& \geq 4 \int_0^{\sqrt{\ell(P_1)}}\!\!\!\!\!\! \!\!\!  r\dd r \left(P_1(1-4r^4)e^{-2 r^2} -(1-P_1) \right)\\
&= F(P_1) = \frac{3 (1-P_1) \left(4 w^2+3\right)}{8 w}+P_1-2 \\
&\text{with} \quad w = w_0\left( \frac{\sqrt{e}}{2}\frac{1-P_1}{P_1}\right) \quad \text{for}
\end{split}
\ee
that is plotted in Fig.~\ref{fig: int neg}. Notably, for $P_1=1$ the bound becomes tight $N_W(\ket{1})= F(1)=\frac{9}{4 \sqrt{e}}-1 \approx 0.36$. To show that the function $F(P_1)$ is convex, one computes
\be
F''(P_1)=\frac{3 w \left(4 w \left(w+2\right)+5\right)}{8 \left(1-P_1\right) P_1^2 \left(w+1\right){}^3},
\ee
which is positive since $w\geq 0$. In the main text, we defined the function $F(P_1)$ continued on the whole interval $P_1\in [0,1]$ by simply setting $F(P_1)=0$ for $P_1\leq \frac{1}{2}.$ Obviously, the continued function remains convex. Furthermore, at $P_1=\frac{1}{2}$ the derivative of the function $F$ is zero $F'(1/2)=0$, and since $F''(P_1) \geq 0$ we can conclude that $F(P_1)$ is non-decreasing on the whole interval.

\section{Finite statistics}
\label{sec:stat}
Consider $n$ independent random variables $X^{(i)}\in[a,b]$ with its mean $\mathds{E}(\bar X )= \mathds{E}(\frac{1}{n}\sum X^{(i)} )$.
The Hoeffding theorem~\cite{Hoeffding1963} gives a simple bound on the deviation of the observed average $\bar X$ after $n$ trails from the expected value $\mathds{E}(\bar X )$
\be
\textrm{P}\Big(\bar X - t \geq \mathds{E}(\bar X) \Big)\leq\exp \left(-\frac{2 n t^2}{(b-a)^2}\right).
\ee
In our case, the observables $X^{(i)}$ takes values in the interval $[-1,3]$ so that $(b-a)^2=16$.

Let us now defined $X^{(i)}$ as the minimum of two variables $X^{(i)}=\min\{X^{(i)}_T, X^{(i)}_R\}$ such that $\bar X = \min \{ \bar X_T, \bar X_R\}$. We have for the probability
\be
\begin{split}
    \textrm{P}\Big(\bar X \geq x \Big) = \textrm{P}\Big(\bar X_T \geq x\,\, \textrm{and} \,\, \bar X_R \geq x\Big) \\
    \leq \textrm{P}\Big(\bar X_T \geq x \Big), \textrm{P}\Big(\bar X_R \geq x \Big).
\end{split}
\ee
We now use $  \bar  P_1 +t \geq x= \min\{\mathds{E}(\bar X_T), \mathds{E}(\bar X_R)\} + t$ such that
\be
 \textrm{P}\Big(\bar X \geq \bar  P_1 +t \Big)\leq \textrm{P}\Big(\bar X \geq \min\{\mathds{E}(\bar X_T), \mathds{E}(\bar X_R)\} + t \Big)
\ee
and consider two cases.

If $\mathds{E}(\bar X_T)\leq \mathds{E}(\bar X_R)$ we  use
\be
\begin{split}
     \textrm{P}&\Big(\bar X \geq \min\{\mathds{E}(\bar X_T), \mathds{E}(\bar X_R)\} + t \Big) \\
     &=  
      \textrm{P}\Big(\bar X \geq \mathds{E}(\bar X_T) + t \Big) \\
     &\leq \textrm{P}\Big(\bar X_T \geq \mathds{E}(\bar X_T) + t\Big) \\
     &\leq  \exp\left(-\frac{2 n t^2}{16}\right).
\end{split}
\ee
Otherwise, we do the same with $\bar X_R$. For both cases we find that 
\be\label{eq: P val temp}
 \textrm{P}\Big(\min \{ \bar X_T, \bar X_R\} - t \geq \bar  P_1  \Big) \leq \exp\left(-\frac{2 n t^2}{16}\right)
\ee
or equivalently
\be
 \textrm{P}\Big(\min \{ \bar X_T, \bar X_R\} - t < \bar P_1  \Big) \geq 1-\exp\left(-\frac{2 n t^2}{16}\right).
\ee
Writing the last expression in the form
\be
\textrm{P}\Big( \hat q_\alpha(\bar X_T, \bar X_R)< \bar P_1  \Big) \geq 1-\alpha
\ee
we find
\be
\hat q_\alpha(\bar X_T, \bar X_R)  =    \min \{ \bar X_T, \bar X_R\} - \sqrt{\frac{16 \log(1/\alpha)}{2 n}},
\ee
the latter being a confidence interval for $\bar P_1$.

\smallskip

\smallskip

For the calibration-dependent benchmark, 
one naturally defines the random variable
\be
Z_T= \begin{cases}
C_1(\hat t, \hat \eta_T) & ( \circ  \bullet) \\
0 & (\circ\circ)  \,\, \textrm{or} \,\, (\bullet\circ)\\
-C_2(\hat t, \hat \eta_T,\hat \eta_R) &  (\bullet \bullet)\\
 \end{cases}
\ee
for the quantity $\hat{P}_1^{T\ast}(\bm p)$ of Eq.~\eqref{parambound}, 
with positive constants $C_1,C_2\geq 0$ giving rise to the confidence interval 
\be
\hat q_\alpha^{T\ast} = \bar Z_T - (C_1(\hat t, \hat \eta_T) +C_2(\hat t, \hat \eta_T ,\hat \eta_R))\sqrt{\frac{\log(1/\alpha)}{2 n}}
\ee
on the average single-photon weight $\bar P_1$. Defining $Z_R$ similarly (with detector's roles exchanged) gives rise to the confidence interval 
\be
\hat q_\alpha^{R\ast} = \bar Z_R - (C_1(\hat r, \hat \eta_R) +C_2(\hat r, \hat \eta_R ,\hat \eta_T))\sqrt{\frac{\log(1/\alpha)}{2 n}},
\ee
by exchanging the roles of the detectors. Both are confidence intervals on $\bar P_1$, that is 
\be
\text{P}\left(\hat q_\alpha^{T(R)\ast}<\bar P_1\right) \geq 1-\alpha.
\ee
It follows that 
\be\begin{split}
\text{P}&\left(\max\{\hat q_\alpha^{T\ast},\hat q_\alpha^{R\ast}\}<\bar P_1\right) \\
&= 
\text{P}\left(\hat q_\alpha^{T\ast}<\bar P_1 \, \& \, \hat q_\alpha^{R\ast}<\bar P_1\right)
    \\
    & =\text{P}\left(\hat q_\alpha^{T\ast}<\bar P_1 \right) \\
    &- \text{P}\left(\hat q_\alpha^{T\ast}< \bar P_1 \, \& \, \hat q_\alpha^{R\ast}\geq\bar P_1\right)\\
    & \geq \text{P}\left(\hat q_\alpha^{T\ast}<\bar P_1 \right)
    - \text{P}\left( \hat q_\alpha^{R\ast}\geq\bar P_1\right)\\
    &\geq 1-2 \alpha,
\end{split}
\ee
hence 
\be
\hat{q}_\alpha^{\ast} =
\max\{ \hat q_{\alpha/2}^{T\ast},\hat q_{\alpha/2}^{R\ast} \}
\ee
is also a confidence interval on $\bar P_1$ with confidence level $1-\alpha$.
\smallskip

Finally, let us discuss the witness of Wigner negativity. First, we label by $Q$ the measured value of $\min\{\bar X_T, \bar X_R\}$ after $n$ measurement rounds and consider the case $Q > 1/2$. Given that $W_{\rho^{(i)}}(0)\leq \frac{1}{\pi}(1-2 P_1^{(i)})$ for each state, for any collection of states that have a positive average Wigner function at the origin $\bar W(0)=\frac{1}{n}\sum_{i=1}^n W_{\rho^{(i)}}(0)\geq 0$, the average single-photon weight is $\bar P_1\leq \frac{1}{2}$.
For such a collection  we thus have
\be\begin{split}
 &\textrm{P}\Big(\min \{ \bar X_T, \bar X_R\}  \geq Q  \Big) \\
 &\leq 
\textrm{P}\Big(\min \{ \bar X_T, \bar X_R\}  - Q \geq \bar P_1 -\frac{1}{2} \Big)\\
&\leq 
\textrm{P}\Big(\min \{ \bar X_T, \bar X_R\}  - \left(Q-\frac{1}{2}\right) \geq \bar P_1 \Big)\\
 &\leq \exp\left(-\frac{2 n \left(Q-\frac{1}{2}\right)^2}{16}\right) 
\end{split}
\ee
by virtue of Eq.~\eqref{eq: P val temp}. In other words, for any collection of states with $\bar W(0) \geq 0$ the probability to get a benchmark value exceeding the observation  $Q>1/2$ is upper bounded by
\be
\text{p-value} \leq  \exp\left(-\frac{2 n \left(\min\{\bar X_T, \bar X_R\}-\frac{1}{2}\right)^2}{16}\right).
\ee
\smallskip

In the calibration-dependent setting with the same argument, the p-value can be obtained analogously. For any value $Q$ and an ensemble of random variables $Z_T^{(1)},\dots,Z_T^{(n)}$ such that $\bar P_1\leq \frac{1}{2}$ one has  
\be\begin{split}
 &\textrm{P}\Big(\bar Z_{T}  \geq Q  \Big) \\
 &\leq 
\textrm{P}\Big(\bar Z_{T}  - Q \geq \bar P_1 -\frac{1}{2} \Big)\\
&\leq 
\textrm{P}\Big(\bar Z_{T}  - \left(Q-\frac{1}{2}\right) \geq \bar P_1 \Big)\\
 &\leq \exp\left(-\frac{2 n \left(Q-\frac{1}{2}\right)^2}{(C_1+C_2)^2}\right).
\end{split}
\ee
Hence, the hypothesis $\bar P_1\leq 1/2$ is associated to the p-value
\be
\text{p-value} \leq \exp\left(-\frac{2 n (\bar Z_T-\frac{1}{2})^2}{(C_1(\hat t, \hat \eta_T) +C_2(\hat t, \hat \eta_T ,\hat \eta_R))^2} \right),
\ee
where $\bar Z_T$ is now the value of $\bar Z_T= \frac{1}{n}\sum_{i=1}^n Z_T^{(i)}$ observed in the experiment. Exchanging the roles of the detectors we also obtain 
\be
\text{p-value} \leq \exp\left(-\frac{2 n (\bar Z_R-\frac{1}{2})^2}{(C_1(\hat r, \hat \eta_R) +C_2(\hat r, \hat \eta_R ,\hat \eta_T))^2} \right).
\ee
Given the observed data $(n_{\circ\circ}, n_{\bullet \circ},n_{\circ \bullet}, n_{\bullet \bullet})$ both bounds are valid statements about all collections of states with $\bar P_1 \leq \frac{1}{2}$. One can then simply choose the most favorable bound, that is 
\be\begin{split}
&\text{p-value} \\
&\leq \min\Big \{\exp\left(-\frac{2 n (\bar Z_T-\frac{1}{2})^2}{(C_1(\hat t, \hat \eta_T) +C_2(\hat t, \hat \eta_T ,\hat \eta_R))^2} \right)\!, \\
    & \qquad \qquad \! \exp\left(-\frac{2 n (\bar Z_R-\frac{1}{2})^2}{(C_1(\hat r, \hat \eta_R) +C_2(\hat r, \hat \eta_R ,\hat \eta_T))^2} \right) \Big \}
\end{split}
\ee
where $\bar Z_T = \frac{1}{n}(C_1(\hat t, \hat \eta_T) n_{\circ \bullet} - C_2(\hat t, \hat \eta_T \hat \eta_R) n_{\bullet\bullet})$ and $\bar Z_R = \frac{1}{n}(C_1(\hat r, \hat \eta_R) n_{ \bullet \circ} - C_2(\hat r, \hat \eta_R \hat \eta_T) n_{\bullet\bullet})$.

\section{Multi-mode product states}
\label{app:multimode}

We now consider multi-mode product states of the form 
\be\label{eq: multimpde product}
\varrho = \bigotimes_k \rho_k,
\ee
where the state of each mode $\rho^{[k]}$ is associated to a probability vector $\bm p^{[k]}$, as defined by the expected values of the operators in Eq.~\eqref{eq: POVM elements}. For multi-mode states, we do not have access to individual values of $\bm p^{[k]}$. Instead, a detector does not click if none of the modes triggers a click. Hence, for the state $\varrho$, the observed probabilities satisfy 
$p_{\circ\circ} = \prod_k p_{\circ\circ}^{[k]}$, 
$p_{\_\circ} = \prod_k p_{\_\circ}^{[k]}$, and
$p_{\circ\_} = \prod_k p_{\circ\_}^{[k]}$.
Denote $P_1^{[k]}=\bra{1}\rho_k\ket{1}$ the single-photon probability for the mode $k$. Under the assumption that the  beam splitter is balanced $p_{\_\circ}^{[k]}=p_{\circ\_}^{[k]}= \sum_n P_n^{[k]}\frac{1}{2^n}$ we will  show in the next section that 
\be
\label{P111}
\widetilde{P}^{T}_1({\bm p})=\frac{1}{2}(12\,  p_{\circ\_}- 9\,  p_{\circ\circ}-4)\leq
\max_k P_1^{[k]}.
\ee
The case of unbalanced beam splitter is analogous  with the single-mode case, we introduce $\widetilde{P}_1^{R}({\bm p})$ which is obtained from the definition of $\widetilde{P}^{T}_1({\bm p})$ (given in Eq.~\eqref{P111}) by replacing $p_{\circ\_}$ by $p_{\_\circ}$. Since either $p_{\circ\_}^{[k]} \geq p_{\_\circ}^{[k]}$ or $p_{\circ\_}^{[k]} \leq p_{\_\circ}^{[k]}$ holds for all modes $k$, the minimum of $p_{\circ\_}^{[k]}$ and $p_{\_\circ}^{[k]}$ is a lower bound on $\sum_n P_n^{[k]}(\lambda)\frac{1}{2^n}$. We deduce that
\be
\max_k P_1^{[k]} \geq
\widetilde {P}_1({\bm p}) = \min\{\widetilde{P}_1^{T}({\bm p}),\widetilde{P}_1^{R}({\bm p})\}.
\ee
Under the assumption that the source produces a multi-mode product state $\varrho$ of the form given in Eq.~\eqref{eq: multimpde product}, 
we thus deduce that there is a mode $k_*$, that can in principle be filtered out, such that the corresponding state $\rho^{[k_*]}$ satisfies $\langle 1| \rho^{[k_*]} | 1\rangle \geq \widetilde {P}_1({\bm p})$.

\subsection{The proof of Eq.~\eqref{P111} }

Consider a multi-mode product state $\varrho = \bigotimes_k \rho_k$ 
with $P_1^{[k]}\leq P$ in each mode. We label  $(p_{\circ\_}^{[k]},p_{\circ \circ}^{[k]})$ the statistics associated to $\rho_k$ and $(p_{\circ\_},p_{\circ \circ})$ the statistics associated to $\rho.$ For a balanced beam splitter, we have  $(p_{\circ\_}^{[k]},p_{\circ \circ}^{[k]})\in \bm Q_P$, with 
\be\begin{split}
\bm Q_P =
\t{Polytope}\Big \{ &(0,0),\left(\frac{1+P}{4},0\right), \\
&\left(\frac{2-P}{2},1-P \right), (1,1)\Big\}. \nonumber
\end{split}
\ee
The probabilities $p_{\circ\_}$ and $p_{\circ\circ}$ satisfy
\be\label{eq: composition}\begin{split}
p_{\circ\_} &= \prod_{k=1}^n p_{\circ\_}^{[k]}, \\
p_{\circ \circ} &=\prod_{k=1}^n p_{\circ \circ}^{[k]}.
\end{split}
\ee
Our first aim is to analyze the possible set of values $\bm Q_P^{\infty}=\{(p_{\circ\_},p_{\circ \circ})\}$ for all $n$ from $1$ to $\infty$ and in particular, to show that $\bm Q_P^{\infty} \subset \bm Q_P$ for $P\geq \frac{1}{2}$. Naturally, we are interested in the extreme points of this set. Eq.~\eqref{eq: composition} is linear in all points $(p_{\circ\_}^{[k]}, p_{\circ\circ}^{[k]})$, hence the extreme points of $\bm Q_P^{\infty}$ are obtained by combining the vertices of $\bm Q_P$. 

Whenever a single vertex $(p_{\circ\_}^{[k]}, p_{\circ\circ}^{[k]})=(0,0)$ appears in the product of Eq.~\eqref{eq: composition}, it results in $(p_{\circ\_},p_{\circ \circ}) =(0,0)$. Similarly, if the vertex $(p_{\circ\_}^{[k]}, p_{\circ\circ}^{[k]})=\left(\frac{1+P}{4},0\right)$ is chosen for at least one mode $k$, $p_{\circ\circ}=0$ and $p_{\circ\_}= \frac{1+P}{4} \prod_{j\neq k} p_{\circ\_}^{[j]} \leq \frac{1+P}{4}$. This means that the point $(p_{\circ\_},p_{\circ \circ}=0)$ remains inside the original polytope $\bm Q_P$. We can thus remember that $(0,0)$ and $\left(\frac{1+P}{4},0\right)$ are points of $\bm Q_P^{\infty}$, but ignore these vertices in the further construction. Analogously, all modes with $(p_{\circ\_}^{[k]}, p_{\circ\circ}^{[k]})=(1,1)$ do not change the value of the product, and we can also ignore this vertex.
 Hence, the only products in Eq.~\eqref{eq: composition} that are potentially not in $\bm Q_P$ are of the form
\be
(p_{\circ\_},p_{\circ \circ})_n=\left(\left(\frac{2-P}{2}\right)^n, \, \left(1-P\right)^n \right)
\ee
for $n\geq 2$. Let us first consider the point $(p_{\circ\_},p_{\circ \circ})_2$. It remains inside  $\bm Q_P$ if and only if
\be\begin{split}
4 p_{\circ\_}-3p_{\circ\circ} -1 &\leq P\\
4 \left(\frac{2-P}{2}\right)^2-3(1-P)^2 -1 &\leq P\\
    P - 2 P^2 &\leq 0 \\
     P &\geq \frac{1}{2}.
\end{split}
\ee
Naturally, if $(p_{\circ\_},p_{\circ \circ})_2 \in \bm Q_P$ the next points $(p_{\circ\_},p_{\circ \circ})_n$ are also in $\bm Q_P$. Therefore, $\bm Q_P^{\infty} \subset \bm Q_P$ for $P\geq \frac{1}{2}$. This means that for any state of the form $\varrho = \bigotimes_k \rho_k$, with $P_1^{[k]} \leq 1/2 \leq P$, $4p_{\circ\_}-3p_{\circ\circ} -1 \leq P.$

\smallskip

\begin{figure}
    \centering
    \includegraphics[width=0.9 \columnwidth]{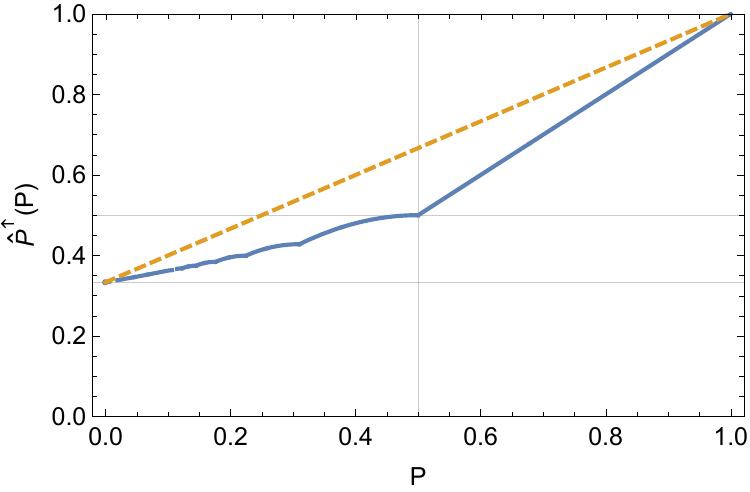}
    \caption{(Full line) Representation of $\hat{P}_1^{T\uparrow}(P)$ as a function of $P$ (full blue line) and an upper-bound (dashed orange line) given by a simple linear function of the form $ \frac{1}{3} +\frac{2}{3} P \geq \hat{P}_1^{T\uparrow}(P)$.}
    \label{fig:multimide bound}
\end{figure}

In order to extend the analysis to any value of $P$, we would need to analyze $\bm Q_P^{\infty}$ for a arbitrary $P$, which is cumbersome.
Instead, we analyze the maximal value that $\hat{P}^T_1({\bm p})$ takes on $\bm Q_P^\infty$. We know that it takes its maximum value on one of the vertices  $(p_{\circ\_},p_{\circ \circ})_n$, and denote these values
\be
\hat{P}^T_1(n)= 4  \left(\frac{2-P}{2}\right)^n-3(1-P)^n - 1,
\ee
for $n\geq 1$. Its maximal value in $\bm Q_P^\infty$ is thus given by
\be\begin{split}
\hat{P}_1^{T\uparrow}(P)&= 
\sup_{\bm p \in \bm Q_P^\infty} \hat{P}^T_1(\bm p ) \\
&= \sup_{n \geq 1}  \hat{P}^T_1(n).
\end{split}
\ee
Let us now look at $\hat{P}^T_1(n) $ 
as a function of a continuous parameter $n\in[0,\infty)$, and compute its derivative
\be
\frac{d}{dn} \hat{P}^T_1(n) = 4 X^n \log(X) - 3 Y^n \log(Y)
\ee
with 
$X =\frac{2-P}{2}$ and $Y =1-P$.
$\hat{P}^T_1(n)$ admits a unique local extremum $\frac{d}{dn} \hat{P}^T_1(n)=0$ at 
\be
n_*=\frac{\log \left(\frac{4 \log (X)}{3 \log (Y)}\right)}{\log (X)-\log (Y)}.
\ee
Furthermore, one easily sees that $\hat{P}^T_1(0)=0$, $\hat{P}^T_1(\infty) = -1$ and $\hat{P}^T_1(1) =P$ and hence $\hat{P}^T_1(n_*)$ is the global maximum of the function.

Next, we recall that $n$ can only take integer values. Thus, the maximal value reads
\be
\hat{P}_1^{T\uparrow}(P)
 =\max\{ \hat{P}^T_1(\lfloor n_*\rfloor), \hat{P}^T_1(\lfloor n_* +1\rfloor).
\ee
It is quite an irregular function, as can be seen in Fig.~\ref{fig:multimide bound}. The boundary value $\hat{P}_1^{T\uparrow}(0)=\lim_{P\to 0} \hat{P}_1^{T\uparrow}(P) =\frac{1}{3}$ can be computed analytically. We show numerically that it is upper-bounded by a simple linear function
\be
\hat{P}^T_1(\bm p)\leq \hat{P}_1^{T\uparrow}(P) \leq \frac{1}{3} +\frac{2}{3} P,
\ee
also can be seen in Fig.~\ref{fig:multimide bound}. By inverting the last inequality we find that 
\be\begin{split}
 \widetilde {P}^T_1(\bm p) &\leq P \qquad \textrm{for}\\
     \widetilde {P}^T_1(\bm p) &= \frac{3 \hat{P}_1^T(\bm p) -1}{2} =\frac{1}{2}(12 p_{\circ \_} - 9 p_{\circ \circ} -4).
\end{split}
\ee
Therefore, for any value $P$ such that $\widetilde {P}_1^T(\bm p)>P$, we can conclude that at least one mode satisfies $\max_k \langle 1|\rho^{[k]}|1\rangle >P$.

\subsection{Finite statistics}
Here, in order to obtain a confidence interval we define an independent random variable
\be
Y_T= \begin{cases}
4 & (\circ \bullet) \\
-1/2 & (\circ\circ)\\
-2 & (\bullet\circ) \,\, \textrm{or} \,\, (\bullet \bullet)\\
 \end{cases},
\ee
with $\mathds{E}(Y_T) = \widetilde {P}_1^{T}(\bm p)$. Similarly, we  define $Y_R$ by  exchanging  the role of the two detectors and get $\mathds{E}(Y_R) = \widetilde {P}_1^{R}(\bm p)$ with $Y_{T(R)}\in[-2,4]$. The same exact analysis as above yields a one-sided confidence interval
\be
\widetilde q_\alpha(\bar Y_T, \bar Y_R)  =    \min \{ \bar Y_T, \bar Y_R\} - \sqrt{\frac{36 \log(1/\alpha)}{2 n}}
\ee
on the quantity $\max_k P_1^{[k]}$ averaged over all states produced by the source. In the general multi-mode case, the same quantity is a one-sided interval on the quantity $\sum_\lambda p(\lambda) \max_k P_1^{[k]}(\lambda)$ averaged over all states produced by the source.

\bibliographystyle{unsrtnat}
\bibliography{references}

\end{document}